\documentclass[12pt]{iopart}
\makeatletter\if@twocolumn\PassOptionsToPackage{switch}{lineno}\else\fi\makeatother
\usepackage{array, ltablex, multirow}
\usepackage{tabulary}
\usepackage[T1]{fontenc}
\usepackage[perpage]{footmisc}
\usepackage{graphicx,subcaption} 
\usepackage[dvipsnames]{xcolor}
\usepackage{tabularx}
\usepackage{hyperref}
\usepackage{stackengine}
\usepackage{marginnote}
\stackMath
\newcommand\tsup[2][2]{%
 \def\useanchorwidth{T}%
  \ifnum#1>1%
    \stackon[-.5pt]{\tsup[\numexpr#1-1\relax]{#2}}{\scriptscriptstyle\sim}%
  \else%
    \stackon[.5pt]{#2}{\scriptscriptstyle\sim}%
  \fi%
}

\usepackage{makecell}
\setcellgapes{5pt}
\makegapedcells

\makeatletter
\expandafter\let\csname equation*\endcsname\relax
\expandafter\let\csname endequation*\endcsname\relax
\usepackage{amsmath}
\expandafter\let\csname subarray\endcsname\relax
\expandafter\let\csname endsubarray\endcsname\relax
\expandafter\let\csname substack\endcsname\relax
\expandafter\let\csname endsubstack\endcsname\relax
\def\[{\relax\ifmmode\@badmath\else
 \begin{trivlist}
 \@beginparpenalty\predisplaypenalty
 \@endparpenalty\postdisplaypenalty
 \item[]\leavevmode
 \hbox to\linewidth\bgroup$ \displaystyle
 \hskip\mathindent\bgroup\fi}
\def\]{\relax\ifmmode \egroup $\hfil \egroup \end{trivlist}\else \@badmath \fi}
\def\equation{\@beginparpenalty\predisplaypenalty
 \@endparpenalty\postdisplaypenalty
\refstepcounter{equation}\trivlist \item[]\leavevmode
 \hbox to\linewidth\bgroup $ \displaystyle
\hskip\mathindent}
\def\endequation{$\hfil \displaywidth\linewidth\@eqnnum\egroup \endtrivlist}
\def\eqnarray{\stepcounter{equation}
   \def\@currentlabel{\p@equation\theequation}%
\global\@eqnswtrue
\global\@eqcnt\z@\tabskip\mathindent\let\\=\@eqncr
\abovedisplayskip\topsep\ifvmode\advance\abovedisplayskip\partopsep\fi
\belowdisplayskip\abovedisplayskip
\belowdisplayshortskip\abovedisplayskip
\abovedisplayshortskip\abovedisplayskip
$$\halign to
\linewidth\bgroup\@eqnsel$\displaystyle\tabskip\z@
 {##{}}$&\global\@eqcnt\@ne $\displaystyle{{}##{}}$\hfil
 &\global\@eqcnt\tw@ $\displaystyle{{}##}$\hfil
 \tabskip\@centering&\llap{##}\tabskip\z@\cr}
\def\endeqnarray{\@@eqncr\egroup
 \global\advance\c@equation\m@ne$$\global\@ignoretrue }
		
\usepackage{iopams,setstack}

\usepackage{url,multirow,morefloats,floatflt,cancel,tfrupee}
\makeatletter

\AtBeginDocument{\@ifpackageloaded{textcomp}{}{\usepackage{textcomp}}}
\makeatother
\usepackage{colortbl}
\usepackage{xcolor}
\usepackage{pifont}
\usepackage[nointegrals]{wasysym}
\makeatletter

\def\mcWidth#1{\csname TY@F#1\endcsname+\tabcolsep}

\def\cAlignHack{\rightskip\@flushglue\leftskip\@flushglue\parindent\z@\parfillskip\z@skip}
\def\rAlignHack{\rightskip\z@skip\leftskip\@flushglue \parindent\z@\parfillskip\z@skip}

\@ifundefined{etal}{}{}

\usepackage{ifxetex}
\ifxetex\else\if@twocolumn\@ifpackageloaded{stfloats}{}{\usepackage{dblfloatfix}}\fi\fi

\AtBeginDocument{
\expandafter\ifx\csname eqalign\endcsname\relax
\def\eqalign#1{\null\vcenter{\def\\{\cr}\openup\jot\m@th
  \ialign{\strut$\displaystyle{##}$\hfil&$\displaystyle{{}##}$\hfil
      \crcr#1\crcr}}\,}
\fi
}

\AtBeginDocument{%
  \@ifpackageloaded{endfloat}%
   {\renewcommand\efloat@iwrite[1]{\immediate\expandafter\protected@write\csname efloat@post#1\endcsname{}}}{\newif\ifefloat@tables}%
}%

\def\BreakURLText#1{\@tfor\brk@tempa:=#1\do{\brk@tempa\hskip0pt}}
\let\lt=<
\let\gt=>
\def\processVert{\ifmmode|\else\textbar\fi}

\@ifundefined{subparagraph}{
\def\subparagraph{\@startsection{paragraph}{5}{2\parindent}{0ex plus 0.1ex minus 0.1ex}%
{0ex}{\normalfont\small\itshape}}%
}{}

\newcommand\role[1]{\unskip}
\newcommand\aucollab[1]{\unskip}
  
\@ifundefined{tsGraphicsScaleX}{\gdef\tsGraphicsScaleX{1}}{}
\@ifundefined{tsGraphicsScaleY}{\gdef\tsGraphicsScaleY{.9}}{}
\def\checkGraphicsWidth{\ifdim\Gin@nat@width>\linewidth
	\tsGraphicsScaleX\linewidth\else\Gin@nat@width\fi}

\def\checkGraphicsHeight{\ifdim\Gin@nat@height>.9\textheight
	\tsGraphicsScaleY\textheight\else\Gin@nat@height\fi}

\def\fixFloatSize#1{}
\let\ts@includegraphics\includegraphics

\def\inlinegraphic[#1]#2{{\edef\@tempa{#1}\edef\baseline@shift{\ifx\@tempa\@empty0\else#1\fi}\edef\tempZ{\the\numexpr(\numexpr(\baseline@shift*\f@size/100))}\protect\raisebox{\tempZ pt}{\ts@includegraphics{#2}}}}

\AtBeginDocument{\def\includegraphics{\@ifnextchar[{\ts@includegraphics}{\ts@includegraphics[width=\checkGraphicsWidth,height=\checkGraphicsHeight,keepaspectratio]}}}

\DeclareMathAlphabet{\mathpzc}{OT1}{pzc}{m}{it}

\def\URL#1#2{\@ifundefined{href}{#2}{\href{#1}{#2}}}

\def\UrlOrds{\do\*\do\-\do\~\do\'\do\"\do\-}%
\g@addto@macro{\UrlBreaks}{\UrlOrds}

\edef\fntEncoding{\f@encoding}

\makeatother

\newif\ifmultipleabstract\multipleabstractfalse%
\usepackage[numbers,square,sort&compress]{natbib}

\usepackage{graphicx}

\begin{document}

\title[]{On the (Non)Hadamard Property of the SJ State in a $1+1$D Causal Diamond}

\author{Yifeng Rocky Zhu$^{1,2}$ and Yasaman K. Yazdi$^2$}
\address{$^1$ DAMTP, Centre for Mathematical Sciences, University of Cambridge, CB3 0WA, UK}
\address{$^2$ Physics Department, Blackett Laboratory, Imperial College London, SW7 2AZ, UK}

\ead{yz792@cantab.ac.uk, ykouchek@imperial.ac.uk}

\vspace{10pt}

\begin{abstract}
The Sorkin-Johnston (SJ) state is a candidate physical vacuum state for a  scalar field in a generic curved spacetime. It has the attractive feature that it is covariantly and uniquely defined in any globally hyperbolic spacetime, often reflecting the underlying symmetries if there are any. A potential drawback of the SJ state is that it does not always satisfy the  Hadamard condition. In this work, we study the extent to which the SJ state in a $1+1$D causal diamond is Hadamard, finding that it is not Hadamard at the boundary. We then study the softened SJ state, which is a slight modification of the original state to make it Hadamard. We use the softened SJ state to investigate whether some peculiar features of entanglement entropy in causal set theory may be linked to its non-Hadamard nature. 
\end{abstract}

\today

\section{Introduction}
Hadamard states are known to have a number of useful properties. In essence, Hadamard states are close to the Minkowski vacuum state and their two-point correlation functions have the same leading singularity structure in the coincidence limit. The desire to have a close relation to the Minkowski vacuum is in part motivated by the equivalence principle. Many states of interest, such as ground and thermal states in stationary spacetimes, are Hadamard  \cite{Sanders:2013}. Hadamard states have also been instrumental  in the perturbative construction of interacting quantum field theories in curved spacetimes \cite{Hollands2001,Brunetti1996,Brunetti_2000,Hollands2002}, the proof of mathematically rigorous quantum energy inequalities \cite{Fewster2000,Fewster2002,Fewster2003}, the regularization of the stress-energy tensor \cite{Wald:1977up, DEWITT1975295,PhysRevD.14.2490,PhysRevD.17.946, Decanini_2008}, 
the description of Unruh-deWitt detectors along general worldlines in curved spacetimes \cite{Junker2002,Louko2006,Fewster2016}, and the analysis of quantum fields on spacetimes with closed timelike curves \cite{Kay1997}. 
However, despite their appeal, there are still some motivations for seeking alternative classes of quantum states that may not be Hadamard.  One is that Hadamard states are non-unique in generic spacetimes, and another that they are not defined in certain settings such as discrete spacetimes.  
Some constraints have been placed on the possibility of alternatives to Hadamard states in the continuum, e.g. in  ultrastatic slab spacetimes in \cite{Fewster_had1}.

Of course, transcending the choice between Hadamard and non-Hadamard states, is the well known challenge of defining a distinguished vacuum state in arbitrary curved spacetimes. In Minkowski spacetime, we are guided by Poincar{\'e} invariance to a unique vacuum state. In static spacetimes,  the timelike and hypersurface-orthogonal Killing vector leads to a distinguished vacuum. However, in general curved spacetimes, there are no symmetries that can be used to single out a  state. Moreover, even the presence of symmetries in some important examples such as cosmological spacetimes (e.g. de Sitter and Friedman-Robertson-Walker spacetimes) is not enough to unambiguously select a preferred state.

Hence, it was a pleasant surprise when the Sorkin-Johnston (SJ) prescription \cite{Sorkin_2011, Johnston_2009, Afshordi_2012, sj_thesis} succeeded to define a distinguished and unique vacuum state for a Gaussian scalar field theory in a large class of globally hyperbolic spacetimes \cite{Afshordi_2012, Fewster_2012}.
The SJ Wightman or two-point function, $W(x,x')=\langle 0|\phi(x)\phi(x')|0\rangle$, is defined in a covariant manner using the Pauli-Jordan or spacetime commutator function, $i\Delta(x,x')=[\phi(x),\phi(x')]$. Since the theory is Gaussian, all of its $n$-point functions are then completely determined in terms of $W$ via Wick's theorem. The SJ state depends only on basic geometric quantities and has been studied in a number of different  continuum spacetimes as well as discrete causal sets \cite{sj_thesis, Afshordi_2012, Aslanbeigi_2013, Buck_2017, Afshordi2_2012, Surya_dssj, Mathur_2019, Avil_n_2014}. This is an advantage over generic Hadamard states which do not necessarily admit a discrete analogue or application. However, one potential drawback of the SJ state is that it is not always Hadamard \cite{Fewster_2012, Fewster_had1, Fewster_had2}. 

This paper aims to study the extent to which the SJ state for a massless scalar field in a causal diamond in $1+1$D Minkowski spacetime is Hadamard. The causal diamond example is arguably the most well-studied case of the SJ state, with a multitude of analytic and numerical results available \cite{sj_thesis, Afshordi2_2012, Saravani_2014}. A gap in the literature on the SJ state in the causal diamond is a thorough and dedicated study of its (non)Hadamard nature, which the present paper intends to fill. More specifically, our work serves two purposes.  First, we review the work on the SJ vacuum in the case of a free massless scalar field in a causal diamond of $1+1$D Minkowski spacetime, closely following \cite{Afshordi_2012}. We also present, for the first time, a detailed study of the portion of the state that does not have a closed-form analytic expression. We then proceed to study both the coincidence limit and non-coincidence behavior of the SJ Wightman function, to probe where it may not be Hadamard. Second, recent results on entanglement entropy in causal set theory indicate the presence of numerous unexpected ultraviolet contributions to the entropy \cite{ee1, ee2, Duffy_2022, Surya_2021}. Since the SJ vacuum is used as the pure state in these calculations, and since the Hadamard condition is often linked to short-distance behavior, we investigate whether the two may be linked. To this end, we make use of \emph{softened} SJ states \cite{Sorkin:2017fcp, Fewster_had2, Wingham:2018pxx} which are known to produce Hadamard (albeit no longer unique) states. We define and discuss these quantities further below.

Our paper is structured as follows. In Section \ref{sec:Hadamard} we discuss the Hadamard condition. In Section \ref{sec: Diamond} we review the SJ prescription and the nature of the SJ state in a causal diamond in $1+1$D, including a calculation determining whether or not the analytically known part of the Wightman function has the Hadamard form. In Section  \ref{sec: Epsilon} we study the remaining (nonanalytic) portion of the Wightman function and through some approximations deduce whether or not it has the Hadamard form. Our results in this section are supplemented with further numerical evidence in the Appendix. Section \ref{sec:Entropy} investigates the softened SJ states and the question of whether or not they yield different results for the entanglement entropy. Finally, we end with a summary of our results and their implications in Section \ref{sec:Summary}.

\section{The Hadamard Condition}
\label{sec:Hadamard}

The expression for a Wightman function with the Hadamard form in $D\geq 2$ spacetime dimensions is

\begin{eqnarray}
W(x,x')=\lim_{\epsilon\rightarrow 0^+}\alpha_D&\left(\frac{U(x,x')}{[\sigma(x,x')+i\epsilon(t-t')]^{\frac{D-2}{2}}}\right.\\
   &\left.+V(x,x')\ln[\frac{\sigma(x,x')+i\epsilon(t-t')}{\ell^2}] +F(x,x')\vphantom{\frac12}\right).
\end{eqnarray}

\noindent 
The expression in $4$ dimensions can be found in \cite{KAY199149}, and the expression above in $D$ dimensions can be inferred (see e.g. \cite{Frob:2017gez}) from the expression for the Feynman propagator in \cite{Decanini_2008} and by replacing the regulator $i\epsilon$ with $i\epsilon(t-t')$.  In the above expression $t, t'$ are time coordinates, $\alpha_D$ is a constant that depends on the dimension, $\ell$ is an arbitrary length scale, $\sigma(x,x')$ is half the square of the geodesic distance between $x$ and $x'$, and $U(x,x')$, $V(x,x')$, and $F(x,x')$ are smooth (or rather, regular) and symmetric functions. The metric signature is $(-,+,+,+)$.

A consequence of the Hadamard condition is that the difference between  the Wightman functions of two Hadamard states is a regular function. Therefore, if we wish to know whether the SJ Wightman function $W_{SJ}$ is Hadamard, we may consider the difference between it and the Minkowski Wightman function $W_M$ (which has the Hadamard form) and check whether the difference is a regular function. This is what we will do below. Since we will study a massless scalar field theory in $1+1$D, the reference Minkowski Wightman function which we will compare to, in Cartesian $(t,\mathbf{x})$ coordinates, is \cite{Afshordi_2012}
\begin{equation}
W_M(t,\mathbf{x};t^\prime,\mathbf{x'}) = - \frac{1}{4 \pi} \textrm{ln} |\sigma| - \frac{i}{4}\textrm{sgn}(t-t')\theta(-\sigma)+H(\mathbf{x},t;\mathbf{x'},t'),
\label{eqn:W_Mink_IR}
\end{equation}
where $2\sigma=(\mathbf{x}-\mathbf{x'})^2-(t-t')^2$, $\theta$ is a step function, and $H(\mathbf{x},t;\mathbf{x'},t')$ is a regular function. Since the massless theory in $1+1$D has an infrared divergence, we must have an IR cutoff in order to obtain the answer \eqref{eqn:W_Mink_IR}; the details of the IR cutoff can be absorbed into $H$. In our setup, the finiteness of the size of the causal diamond will serve as the IR cutoff.

\section{The SJ Vacuum in a Causal Diamond}
\label{sec: Diamond}
The SJ prescription \cite{Sorkin_2011, Johnston_2009, Afshordi_2012, sj_thesis} was initially proposed by Sorkin and Johnston as a means to study quantum fields on causal sets, where the notion of a state on a hypersurface has no physical reality. Shortly after its introduction, its application to continuum spacetimes was also realized. The starting point is the retarded Green function, $G_R$, from which one obtains the spacetime commutator, as we see below. Due to this, the prescription yields a unique state only in globally hyperbolic spacetimes, where a unique retarded Green function exists. The causal diamond in $1+1$D Minkowski spacetime is one of the simplest examples of a globally hyperbolic spacetime. Below, we review and extend what is known about the nature of the SJ state in the $1+1$D Minkowski diamond.

\subsection{The Sorkin-Johnston Prescription}

Let us begin by reviewing the SJ prescription. The retarded Green function of a scalar field can be obtained from the Klein-Gordon equation, as it satisfies
\begin{equation}
(\Box+m^{2})G_R(x,x')=-\frac{\delta^{D}(x-x')}{\sqrt[]{-g}},
\end{equation}
where $G_R(x,x')$ is only non-zero if $x'$ causally precedes $x$.

From the retarded Green function, we can define the Pauli-Jordan function $\Delta(x,x^\prime)$ as
\begin{equation}
\Delta(x,x') = G_R(x,x') - G_R(x',x).
\end{equation}
The commutator $i\Delta$ can be made into an integral kernel acting on $L^2(\mathcal{M})$ functions over a spacetime $\mathcal{M}$ as
\begin{equation}
\tilde{v}(x)=\int_\mathcal{M}(i\Delta)(x,x')v(x')\,dV_{x'},
\end{equation}
$v$ being an arbitrary square integrable function on the spacetime and $dV_{x'}$ being the volume measure with respect to the spacetime metric. Since the commutator $i\Delta$ is antisymmetric, its eigenfunctions come in complex conjugate pairs $T^\pm$ with real corresponding eigenvalues $\pm \lambda$. Furthermore, the spectral decomposition of $i\Delta(x,x^\prime)$ implies that we may write it as
\begin{equation}
i \Delta(x,x^\prime) = \sum_{q} \lambda_q T_q^+(x) T_q^+(x^\prime)^* - \sum_{q} \lambda_q T_q^-(x) T_q^-(x^\prime)^*.
\label{eqn:kernel}
\end{equation}
The SJ two-point correlation function $W_{SJ}(x,x^\prime)$ is then defined by restricting \eqref{eqn:kernel} to its positive part
\begin{equation}
W_{SJ}(x,x^\prime) \equiv \sum_{q} \lambda_q T_q^+(x) T_q^+(x^\prime)^*.
\label{eqn:W_SJ}
\end{equation}

\subsection{$W_{SJ}$ in the Diamond} \label{sec:wsj_diamond}

We limit ourselves to the massless theory in $1+1$D. The retarded Green function for this theory is

\begin{equation}
G_R(x,x^\prime) = \frac{1}{2}\theta\left((t-t')^2-(\mathbf{x}-\mathbf{x'})^2\right)\theta(t-t').
\label{eqn:Green}
\end{equation}
We will hereafter adopt the lightcone coordinates  $u = (t+\mathbf{x})/ \sqrt{2}$ and $v = (t-\mathbf{x})/ \sqrt{2}$ for convenience. In a causal diamond of side length $2L$ and centered at $u = v= 0$ (see e.g. Figure \ref{fig:diamond}), the domains of these coordinates are $u\in[-L,L]$ and $v\in[-L,L]$.

The positive eigenfunctions $T_k^+$
that satisfy $i \int_\diamond \Delta(x,x') T_k^+(x') dx = \lambda_k T_k^+(x)$ are given by the two sets of functions \cite{Johnston_PhD}
\begin{equation}
\eqalign{
f_k(u,v):=& e^{-iku} - e^{-ikv} , ~~ \qquad \qquad \qquad \textrm{with}~k = \frac{n \pi }{L},~ n = 1,2,... \\
g_k(u,v):=& e^{-iku} + e^{-ikv} - 2\textrm{cos}(kL), \qquad \textrm{with}~k \in \mathcal{K},
}
\end{equation}
where $\mathcal{K} = \{ k \in \mathbb{R} ~|~ \textrm{tan}(kL) = 2kL ~~\textrm{and}~~ k > 0 \}.$
We can then go on to construct the SJ Wightman function by summing over the positive eigenvalues and their corresponding eigenfunctions, as in \eqref{eqn:W_SJ}. Since there are two families of eigenfunctions, ${f_k}$ and ${g_k}$, let us consider the sum over each set separately

\begin{equation}
W_{SJ,L}(u,v;u^\prime,v^\prime) = S_1 + S_2,
\label{eqn:W_SJL}
\end{equation}
where 

\begin{equation}
\eqalign{
S_1 &= \sum_{n=1}^\infty \frac{L^2}{n\pi}\frac{1}{\|f_k \|^2}f_k(u,v)f_k^*(u^\prime,v^\prime)   \\
&= \frac{1}{8\pi}\bigg\{ 
-\textrm{ln}\bigg[1 - e^{-\frac{i\pi(u-u^\prime)}{L}} \bigg] -
\textrm{ln}\bigg[1 - e^{-\frac{i\pi(v-v^\prime)}{L}} \bigg]\label{eq:s1sum} \\
& \quad \quad \quad \quad \quad \quad
+ \textrm{ln}\bigg[1 - e^{-\frac{i\pi(u-v^\prime)}{L}} \bigg] +
\textrm{ln}\bigg[1 - e^{-\frac{i\pi(v-u^\prime)}{L}} \bigg] \bigg\}},
\end{equation}
and
\begin{equation}
\eqalign{
S_2 
&=  \sum_{k \in \mathcal{K}} \frac{L}{k}\frac{1}{\|g_k \|^2}g_k(u,v)g_k^*(u^\prime,v^\prime) \\
&= \sum_{k \in  \mathcal{K}}\frac{[e^{-iku} + e^{-ikv} -2\textrm{cos}(kL) ]
[e^{iku^\prime} + e^{ikv^\prime} -2\textrm{cos}(kL)]
}{kL[8-16\textrm{cos}^2(kL)]}.
}
\label{eqn:S2}
\end{equation}
Note that $S_2$ does not have a closed form expression, due to the transcendental equation involved in finding the wavenumbers. However, the positive roots $x_n$ of the transcendental equation $\textrm{tan}(x) = 2x$ rapidly approach $x_n =\frac{(2n-1)\pi}{2}$ as $n \rightarrow \infty$ (with $n \in \mathbb{N}$). Therefore, we can approximate the wavenumbers as \textit{k}$_n^{(0)}$ = $\frac{(2n-1)\pi}{2L}$ and obtain an approximate analytic contribution to the sum in \eqref{eqn:S2}. We will call $S_2^{(0)}$ this approximate analytical portion obtained from $k_n=\textit{k}_n^{(0)}$ = $\frac{(2n-1)\pi}{2L}$. This approximation   introduces an error in our calculations, which following \cite{Afshordi2_2012} we will denote as $\epsilon$. $S_2$ is now given by
\begin{equation}
\eqalign{
S_2 &= S_2^{(0)} + \epsilon(u,v;u^\prime,v^\prime) \\
&= \frac{1}{4\pi} \bigg\{ 
\textrm{tanh}^{-1} \bigg[ e^{-\frac{i\pi(u-u^\prime)}{2L}}\bigg] +
\textrm{tanh}^{-1} \bigg[ e^{-\frac{i\pi(v-v^\prime)}{2L}}\bigg] \\
& \quad 
+ \textrm{tanh}^{-1} \bigg[ e^{-\frac{i\pi(u-v^\prime)}{2L}}\bigg] +
\textrm{tanh}^{-1} \bigg[ e^{-\frac{i\pi(v-u^\prime)}{2L}}\bigg]
\bigg\} + \epsilon(u,v;u^\prime,v^\prime)
},
\label{eq:s2approx}
\end{equation}
 where $S_2^{(0)}$ is the analytical part, and $\epsilon$ as mentioned is the error from the approximation. Therefore, adding \eqref{eq:s2approx} and \eqref{eq:s1sum} and simplifying, we may rewrite \eqref{eqn:W_SJL} as
\begin{equation} \label{eq:s2approx_simplified}
\eqalign{
W_{SJ,L} 
&= \frac{1}{4\pi} \bigg\{ 
- \textrm{ln}\bigg[1 - e^{-\frac{i\pi(u-u^\prime)}{2L}} \bigg] 
- \textrm{ln}\bigg[1 - e^{-\frac{i\pi(v-v^\prime)}{2L}} \bigg] \\ 
&  \quad 
+ \textrm{ln}\bigg[1 + e^{-\frac{i\pi(u-v^\prime)}{2L}} \bigg] 
+ \textrm{ln}\bigg[1 + e^{-\frac{i\pi(v-u^\prime)}{2L}} \bigg] 
\bigg\} + \epsilon(u,v;u^\prime,v^\prime).
}
\end{equation}
Taking the coincidence limit ($u \rightarrow u^\prime$ and  $v \rightarrow v^\prime$), we obtain 
\begin{eqnarray}
\eqalign{
W_{SJ,L} &\approx 
-\frac{\textrm{ln}|\Delta u \Delta v | }{4\pi}
- \frac{1}{2\pi} \textrm{ln} \bigg( \frac{\pi}{L} \bigg) +
\frac{\textrm{ln}(4)}{4\pi}
-
\frac{i}{4}\textrm{sgn}(\Delta u + \Delta v)
\theta(\Delta u \Delta v)
\\&  \quad \quad
+ \frac{1}{4\pi} \textrm{ln}\bigg[1 + e^{-\frac{i\pi(u-v^\prime)}{2L}} \bigg]
+ 
\frac{1}{4\pi} \textrm{ln}\bigg[1 + e^{-\frac{i\pi(v-u^\prime)}{2L}} \bigg]
+\epsilon(u,v;u^\prime,v^\prime),
}
\label{eqn:Wgeneral}
\end{eqnarray}

\noindent where $\Delta u\equiv u-u'$ and $\Delta v\equiv v-v'$. The first line of \eqref{eqn:Wgeneral} agrees with the IR regulated Minkowski two-point function (\ref{eqn:W_Mink_IR}) and therefore has the Hadamard form. In order for \eqref{eqn:Wgeneral} to be Hadamard, the expression in its second line must be a smooth (or rather, $C^2$) function. However, the two logarithmic terms in the second line of both \eqref{eq:s2approx_simplified} and \eqref{eqn:Wgeneral} (which is unaltered by the coincidence limit) diverge when 
\begin{equation} \label{eqn:divergence_conditions}
u - v^\prime = \pm 2L \quad \text{or} \quad v - u^\prime = \pm 2L,
\end{equation}

\begin{figure}[t]
 \begin{subfigure}{0.4\textwidth}
     \includegraphics[width=\textwidth]{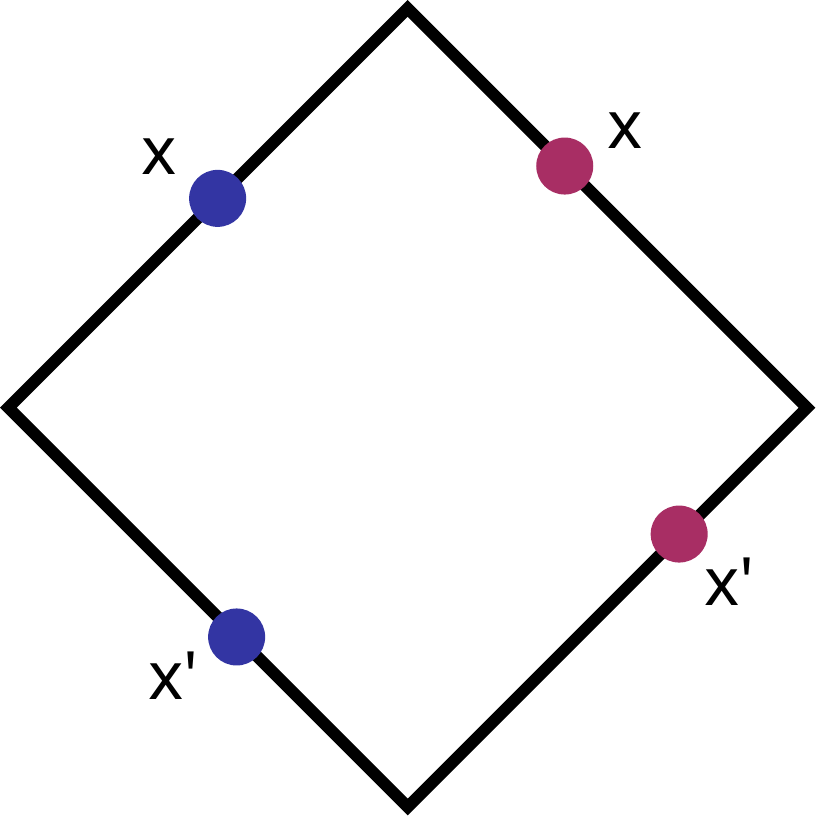}
     \caption{$x$ on the upper edges}
     \label{fig:x_above}
 \end{subfigure}
 \hfill
 \begin{subfigure}{0.4\textwidth}
     \includegraphics[width=\textwidth]{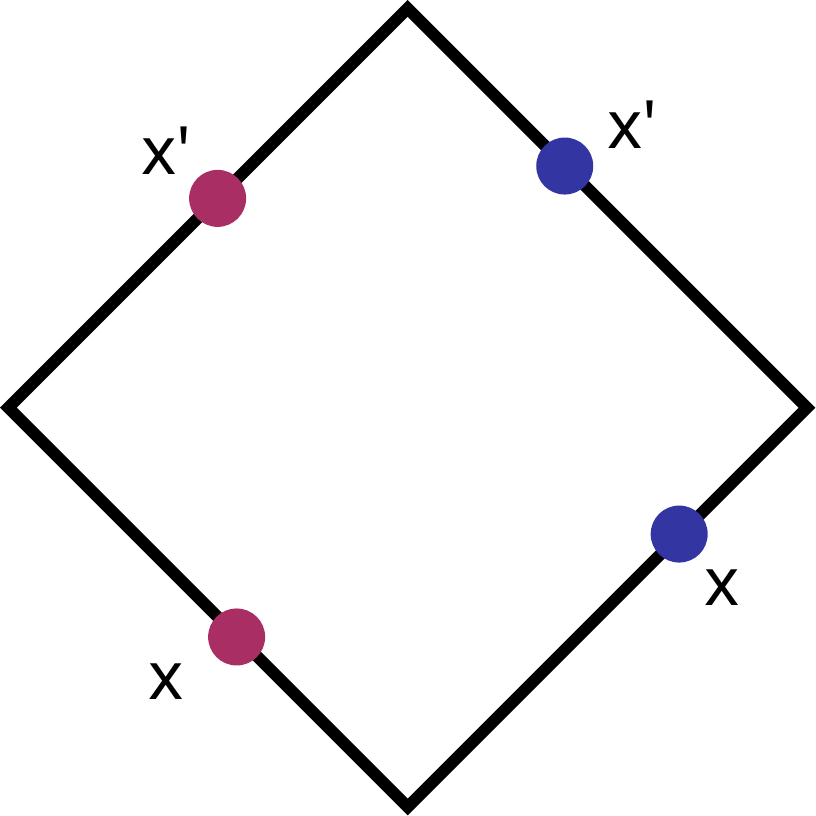}
     \caption{$x$ on the lower edges}
     \label{fig:x_below}
 \end{subfigure}
 \caption{Pairs of points $x$ and $x'$ that satisfy the divergence conditions in \eqref{eqn:divergence_conditions}.  These conditions are satisfied when both points are on the two neighbouring left or right edges of the boundary of the diamond. The two pairs of red points satisfy the condition $u-v'=\pm 2L$, and the two pairs of blue points satisfy the condition $v-u'=\pm2L$.  In the coincidence limit when both points lie on either the left or right corner, both conditions are satisfied.  }
 \label{fig:nonlocal_divergences}
\end{figure}

\noindent
respectively.  These conditions are satisfied when both inputs $x$ and $x'$ of $W_{SJ,L}$ lie on either the left two adjacent edges of the diamond, or on the right two edges.\footnote{Note that both the $f$ and $g$ modes contribute to the non-Hadamard logarithmic divergences discussed here, as the $u,v'$ and $v,u'$ terms diverge in both equations \eqref{eq:s1sum} and \eqref{eq:s2approx} when the respective conditions of \eqref{eqn:divergence_conditions} are satisfied.  The divergence of the relevant term from the $g$-mode follows from the fact that $\tanh^{-1}(x)=\frac{1}{2}\ln(1+x)-\frac{1}{2}\ln(1-x)$.}  The possibilities are enumerated in Figure \ref{fig:nonlocal_divergences}. In these cases, when the points do not coincide, exactly
one of the logarithmic terms in the second line of \eqref{eq:s2approx_simplified} will diverge.  In the coincidence limit, when both $x$ and $x'$ lie on either the left or right corner of the diamond, both conditions in \eqref{eqn:divergence_conditions} are satisfied and both log-terms in the second line of \eqref{eqn:Wgeneral} will diverge.  
Therefore, we find a potential deviation from the Hadamard form on the boundary of the diamonds, on the condition that the $\epsilon$ term does not have compensating divergences (which is checked in the next section).  The failure of the SJ state to be Hadamard on the left and right corners of the diamond was already known in the literature, and was previously observed in \cite{Fewster_had1}.   Note that derivatives of these two log terms do not lead to divergences in new locations.

What remains to be checked is the nature of the $\epsilon$ term. As the treatment of this term is delicate and is entirely new work, we devote the next section to it.

\section{The Epsilon Term}
\label{sec: Epsilon}
For the Hadamard condition, we would like to know if  $\epsilon$ is a $C^2$ function. In what follows, our strategy will be to approximate $\epsilon$ and its first two derivatives analytically, by treating $S_2$ as a Taylor expansion around $k_n=k_n^{(0)}$. We also use a Taylor expansion of the transcendental equation $\textrm{tan}(kL) = 2kL$ to approximate the correction to $k_n^{(0)}$. This will allow us to obtain analytic results for the behavior of $\epsilon$. However, by considering these two Taylor series which are not independent, we lose some accuracy in the expansion. In the Appendix we use more accurate numerical results to then further support our claims. 
\subsection{$\epsilon(u,v;u^\prime,v^\prime)$}
Recall that the analytic portion of \eqref{eqn:S2} was obtained by making the approximation $k_n\approx k_n^{(0)}$ = $\frac{(2n-1)\pi}{2L}$. 
Let us consider the contribution to $\epsilon$ from each $n^{th}$ root (i.e. from each term $S_{2,n}(k_n,u,v,u^\prime,v^\prime)$ in the sum \eqref{eqn:S2}):
\begin{equation}
\epsilon_n(u,v;u^\prime,v^\prime) \equiv S_{2,n}(k_n,u,v;u^\prime,v^\prime) - S_{2,n}( k_n^{(0)},u,v;u^\prime,v^\prime).
\label{eqn:epsilon_exact_n}
\end{equation}
\noindent
Then $\epsilon$ is given by
\begin{equation}
\epsilon = S_2 - S_2^{(0)} = \sum_{n=1}^{\infty} \epsilon_n.
\label{eqn:epsilon_exact}
\end{equation}
\noindent
However, the infinite sum in \eqref{eqn:epsilon_exact} does not evaluate to a closed form analytic expression (as we know from \eqref{eqn:S2}).
We will evaluate an approximate analytic expression for it. To approach this, we begin by estimating the correction $\delta_n$ to the approximation $k_n^{(0)}$ for $k_n$. Expanding the transcendental equation tan(\textit{kL}) = 2\textit{kL} around  \textit{k}$_n^{(0)}L$ = $\frac{(2n-1)\pi}{2}$, we have 
\begin{equation}
\eqalign{
F(k) &= \textrm{tan}(kL) - 2kL = 0 \\
F(k_n^{(0)}L+\delta_n) &=  \textrm{tan}(k_n^{(0)}L+\delta_n) - 2(k_n^{(0)}L + \delta_n)  \\
&= \frac{-\textrm{cos}(\delta_n)}{\textrm{sin}(\delta_n)} - 2k_n^{(0)}L - 2\delta_n=0,
}
\end{equation}
\noindent
hence
\begin{equation}
\eqalign{
\textrm{cot}(\delta_n) + 2k_n^{(0)}L + 2\delta_n &= 0  \\
\frac{1}{\delta_n} - \frac{\delta_n}{3}  - \frac{\delta_n^3}{45} - \frac{2\delta_n^5}{945}
+ O(\delta_n^7) + 2k_n^{(0)}L + 2\delta_n &= 0  \\
\rightarrow k_n^{(0)}L = -\frac{1}{2\delta_n} - \frac{5\delta_n}{6} + \frac{\delta_n^3}{90} + \frac{\delta_n^5}{945} +  O(\delta_n^7)
}
\end{equation}
\noindent
Then, using series reversion we obtain 
\begin{equation}
\delta_n = -\frac{1}{2k_n^{(0)}L} - \frac{5}{24(k_n^{(0)}L)^3}-\frac{83}{480(k_n^{(0)}L)^5} - O\bigg(\frac{1}{(k_n^{(0)}L)^7}\bigg).
\label{eqn:delta}
\end{equation}
We start by correcting $k_n^{(0)}$ with the leading term of $\delta_n$ only. Expressing $S_{2,n}(k_n,u,v,u^\prime,v^\prime)$ as a series around $k_n^{(0)}$ yields
\begin{equation}
\eqalign{
S_{2,n}(k_n,u,v,u^\prime,v^\prime) \approx & \,
S_{2,n}(k_n^{(0)},u,v,u^\prime,v^\prime)  \\
& +  \sum_{j=1}^{\infty}
\frac{1}{j!} \bigg(\frac{-1}{2k_n^{(0)}L} \bigg)^j
\left. \frac{\partial^j S_{2,n}(k_n,u,v;u^\prime,v^\prime)}{\partial k_n^j} \right|_{k_n=k_n^{(0)}},
}
\label{eqn:S2_1st_order}
\end{equation}
\noindent
where the $j^{th}$ order expansion of the approximate error, $\epsilon^{(j)}$, is given by
\begin{equation}
\eqalign{
\epsilon^{(j)} &= \,
\sum_{n=1}^\infty \epsilon_n^{(j)} 
= \,  \sum_{n=1}^\infty
\frac{1}{j!} \bigg(\frac{-1}{2k_n^{(0)}L} \bigg)^j
\left. \frac{\partial^j S_{2,n}(k_n,u,v;u^\prime,v^\prime)}{\partial k_n^j} \right|_{k_n=k_n^{(0)}}, \\
}
\label{eqn:epsilon_j}
\end{equation}

\begin{figure}[t]
 \begin{subfigure}{0.5\textwidth}
     \includegraphics[width=\textwidth]{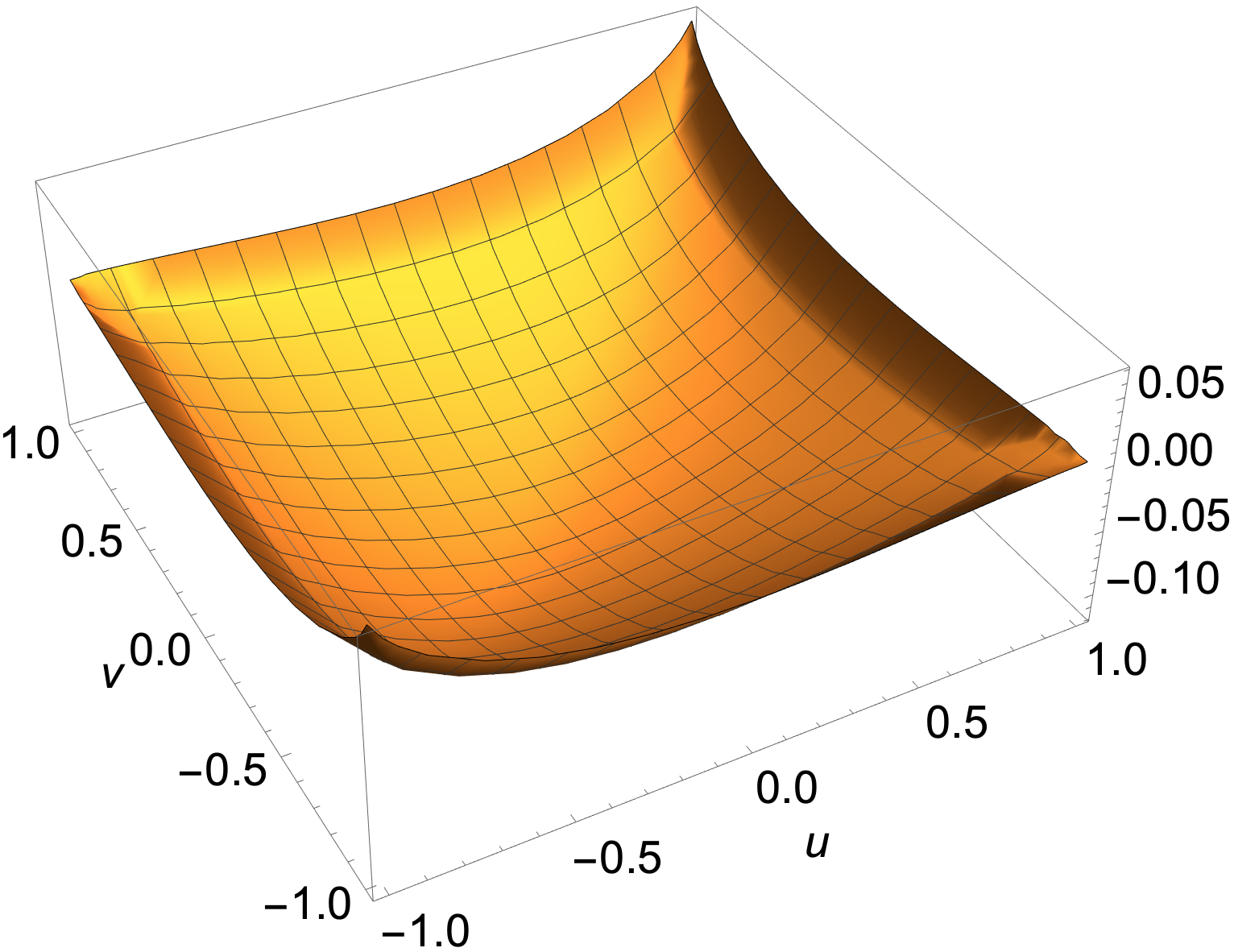}
     \caption{$\epsilon^{(1)}$}
     \label{fig:1st_order}
 \end{subfigure}
 \hfill
 \begin{subfigure}{0.5\textwidth}
     \includegraphics[width=\textwidth]{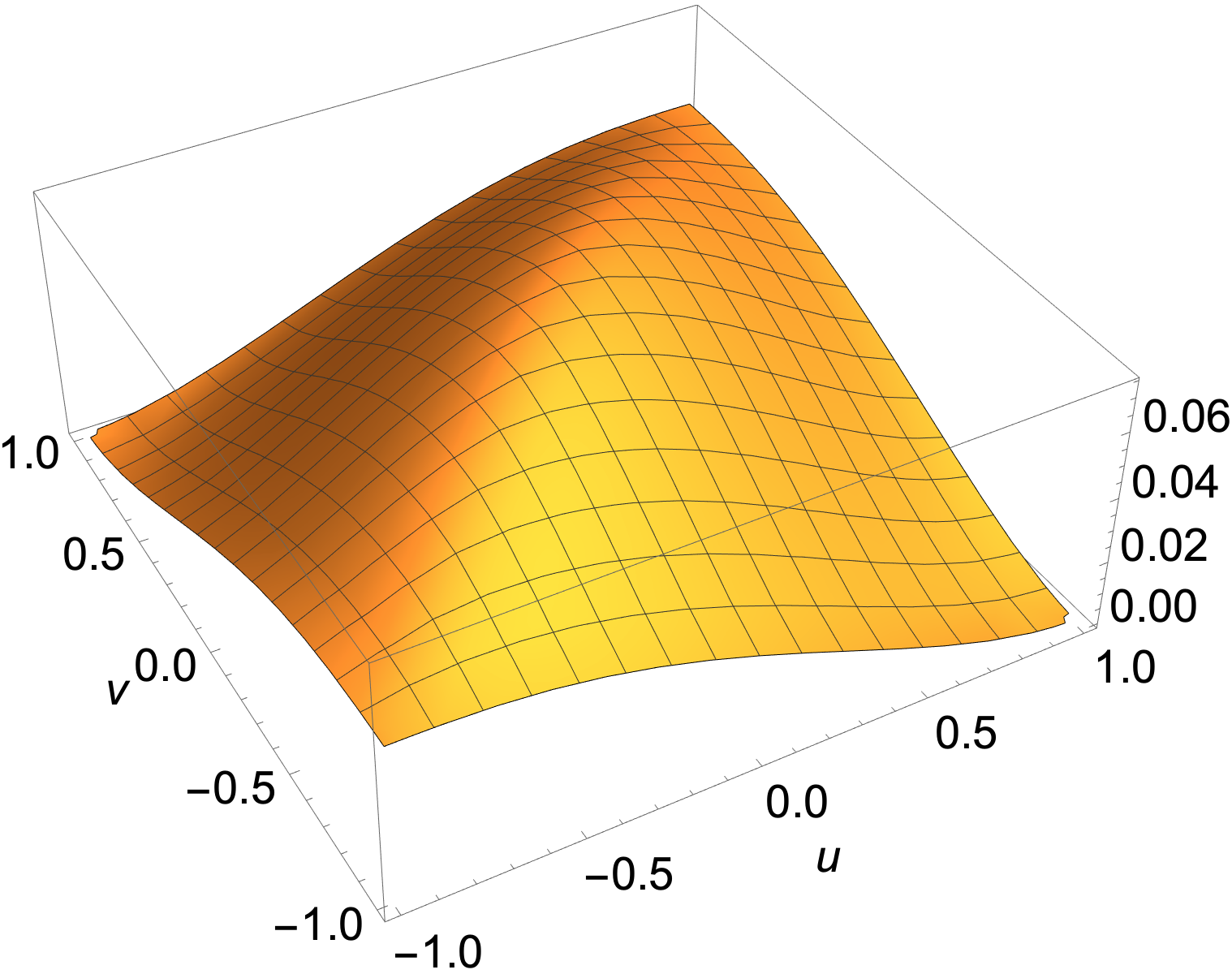}
     \caption{$\epsilon^{(2)}$}
     \label{fig:2nd_order}
 \end{subfigure}
 \caption{ The approximate errors at the  $1^{st}$ and $2^{nd}$ orders of the series, $\epsilon^{(1)}$ and $\epsilon^{(2)}$ in the coincidence limit $u'\rightarrow u$, $v'\rightarrow v$. The plots demonstrate that these functions are finite in the full domain of the causal diamond.}  
 \label{fig:epsilon_1st2nd}
\end{figure}
\noindent
and where $ \epsilon_n^{(j)} $ denotes the approximate error from the $n^{th}$ root at the $j^{th}$ order of this series. Now,
\begin{equation}
\epsilon \approx \sum_{j=1}^{\infty} \epsilon^{(j)}.
\label{eqn:epsilon_approx}
\end{equation}
We can expect the main contribution to the (approximated) error to come from the first few terms in the expansion \eqref{eqn:epsilon_approx}.  This is based on the numerical observation that the absolute value of the real part of the individual terms in the $n,j$ expansion in \eqref{eqn:epsilon_j}, i.e.
\begin{equation*}
\left|Re\left[\frac{1}{j!} \bigg(\frac{-1}{2k_n^{(0)}L} \bigg)^j
\left. \frac{\partial^j S_{2,n}(k_n,u,v;u^\prime,v^\prime)} {\partial k_n^j} \right|_{k_n=k_n^{(0)}}\right]\right|,
\end{equation*}
exhibit a faster than inverse power law decay to zero as a function of $j$, when evaluated for fixed $n$ at random coordinates $(u,v)$, $(u',v')$, in the interior as well as on the boundary of the diamond.  Note that this observation holds for both the coincidence limit as well as non-coincidence input points.  Due to the fast decay of these terms for all $n$, we can reasonably expect the first few terms $\epsilon^{(j)}=\sum_{n=1}^{\infty}\epsilon^{(j)}_n$ in \eqref{eqn:epsilon_approx}, to make up the main contribution to the approximated error term $\epsilon$.  Furthermore, the coefficients of the various powers of $n$ in the series expansion \eqref{eqn:delta} for $\delta_n$ tend to zero faster than an inverse power law as well.  In other words, the series expansion for $\delta_n$ converges quickly, and approximating $\delta_n$ by the leading term in \eqref{eqn:delta},  i.e. $\frac{-1}{2k_n^{(0)}L}$, should therefore provide a reasonable approximation to the full error term $\epsilon$ when expanding $S_{2,n}$ in \eqref{eqn:S2_1st_order}.  \ref{app:NumericalEpsilon} gives numerical evidence that (in the coincidence limit) the difference between the full error $\epsilon$, and the part obtained when approximating $\delta_n$ in this way, is bounded and does not miss further divergences in $\epsilon$.

Based on the above discussion, we therefore focus on the leading terms $\epsilon^{(1)}$ and $\epsilon^{(2)}$. The sum \eqref{eqn:epsilon_j} can be evaluated analytically for both $j=1$ and $j=2$. We do not show the resulting expressions here, as they are complicated and not particularly illuminating to see explicitly. The coincidence limits $u\rightarrow u'$ and $v\rightarrow v'$ can also be evaluated, meaning that these limits are well defined and finite. Again, we do not show the explicit expressions, as they are lengthy and not very useful to see. Instead, in Figure \ref{fig:epsilon_1st2nd} we show plots of the coincidence limits of $\epsilon^{(1)}(u,v;u,v)$ and $\epsilon^{(2)}(u,v;u,v)$, with  \textit{L} = 1. As can be seen in the figure, $\epsilon^{(1)}$ and $\epsilon^{(2)}$ are finite everywhere in the diamond.

\begin{figure}[h]
 \begin{subfigure}{0.45\textwidth}
     \includegraphics[width=\textwidth]{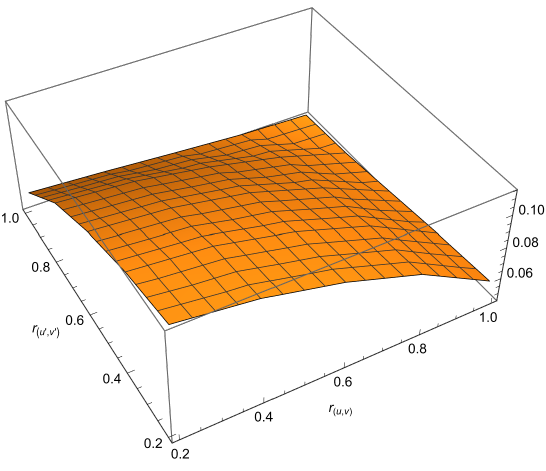}
     \caption{$\left.\max(|Re[\epsilon^{(1)}]|)\right|_{(r_{(u,v)},r_{(u',v')})}$}
     \label{fig:eps1_nonlocal}
 \end{subfigure}
 \hfill
 \begin{subfigure}{0.45\textwidth}
     \includegraphics[width=\textwidth]{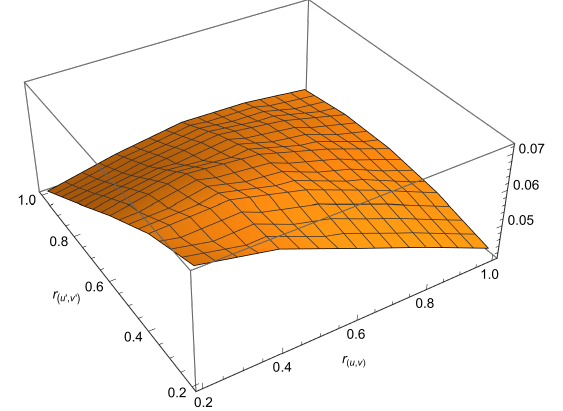}
     \caption{$\left.\max(|Re[\epsilon^{(2)}]|)\right|_{(r_{(u,v)},r_{(u',v')})}$}
     \label{fig:eps2_nonlocal}
 \end{subfigure}
 
 \medskip
 \begin{subfigure}{\textwidth}
 \hspace{0.24\textwidth}
     \includegraphics[width=0.6\textwidth]{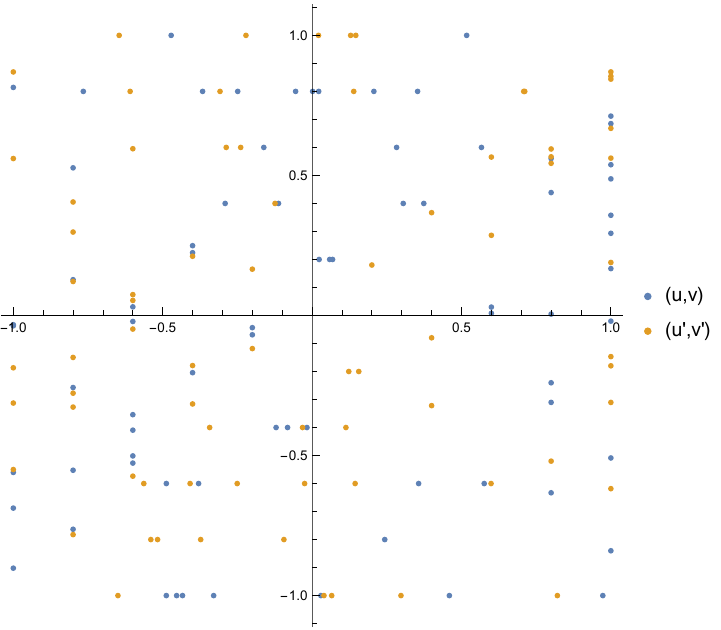}
     \caption{Sampled points $x$ and $x'$.}
     \label{fig:nonlocal_points}
 \end{subfigure}
 \caption{The top two plots show the maximum absolute value of the real part of $\epsilon^{(1)}$ and $\epsilon^{(2)}$, when evaluated on pairs of points at different ``radii'' $r_{(u,v)}$ and $r_{(u',v')}$, corresponding to the $(u,v)$ and $(u',v')$ coordinates shown in the bottom plot.} 
 \label{fig:eps_nonlocal}
\end{figure}

In the case where the points $x$ and $x'$ do not coincide, it is harder to establish that $\epsilon^{(1)}$ and $\epsilon^{(2)}$ are still well behaved everywhere, since their expressions are quite complicated and cannot be visualized as easily, being dependent on 4 variables.  Instead, we evaluate the functions on the boundaries of different subdiamonds, with ``radii'' $r=0.2L$, $0.4L$, $0.6L$, $0.8L$ and $L$, where a radius $r$ defines the edges $\{u\in [-r,r], v=\pm r\}$, or $\{u=\pm r, v\in[-r,r]\}$.  We then sample two sets of random points lying on these boundaries, one for the unprimed points and one for the primed points (as shown in Figure \ref{fig:nonlocal_points}), and evaluate the functions for each pair of points in the two sets. We then determine the \emph{maximum absolute real} part of the function out of all pairs of points coming from two radii and plot those values as the $z$ value of the coordinates $(r_{(u,v)},r_{(u',v')},z)$ in Figures \ref{fig:eps1_nonlocal} and \ref{fig:eps2_nonlocal} for $\epsilon^{(1)}$ and $\epsilon^{(2)}$, respectively.  As can be seen, these two functions are smooth and finite everywhere within and on the boundary of the diamond.  While not a proof of the absence of singularities for special choices of points $x$ and $x'$, the plots suggest that the two functions remain finite.

Given the finiteness of $\epsilon^{(1)}$ and $\epsilon^{(2)}$ in both the coincidence and nonlocal cases, we have a good indication that  $\epsilon$ is at least $C^0$ everywhere. Hence we observe no evidence of additional non-Hadamard behavior from $\epsilon$ (or, any cancelling divergences of the non-Hadamard logarithmic divergences noted in Section \ref{sec:wsj_diamond}, for that matter). \\

\subsection{First Order Derivatives}
\begin{figure}[h]
\begin{subfigure}{0.45\textwidth}
\includegraphics[width=\textwidth]{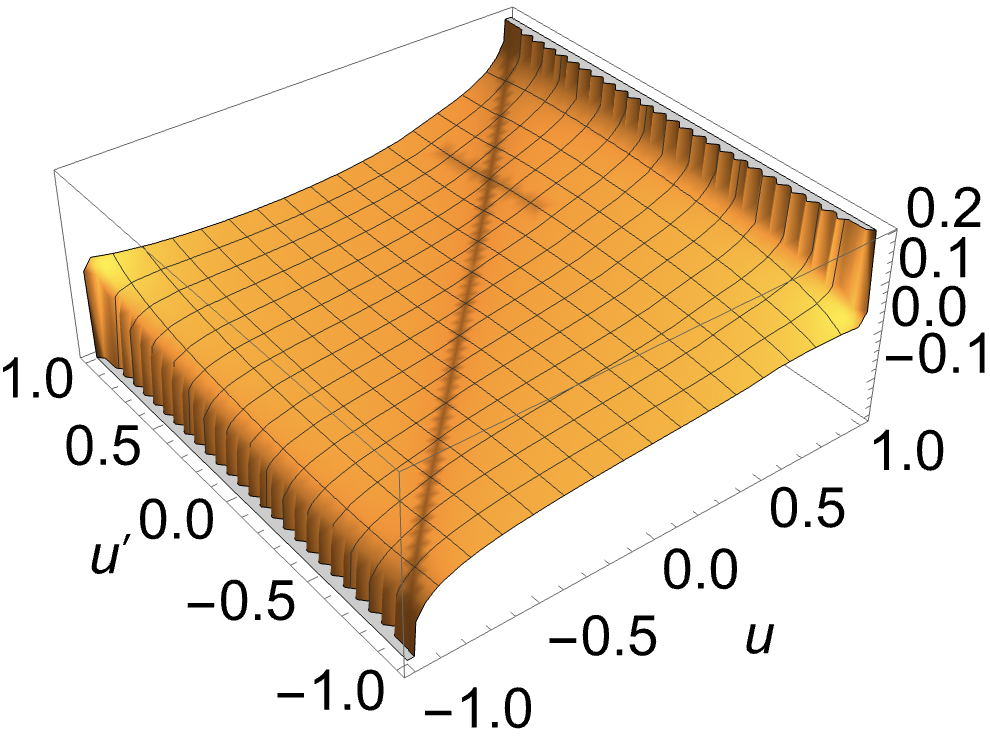}
\caption{Re[$\epsilon^{(1)}_{,u}$]}
\end{subfigure}
\hfill
\begin{subfigure}{0.45\textwidth}
\includegraphics[width=\textwidth]{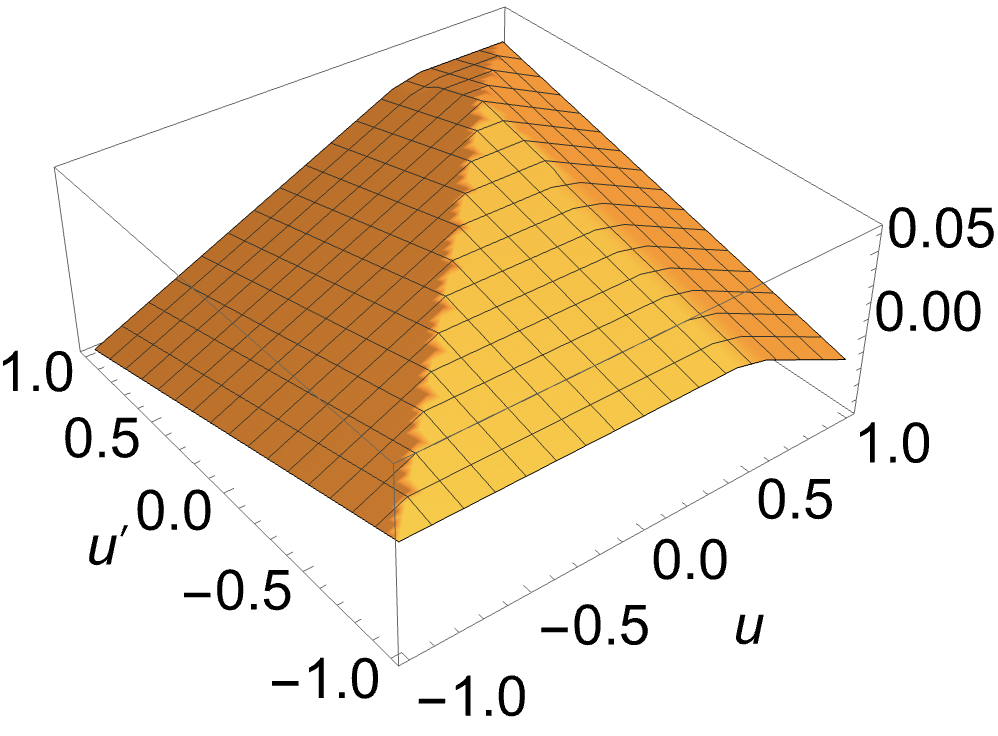}   
\caption{Im[$\epsilon^{(1)}_{,u}$]}
\end{subfigure}

\medskip

\begin{subfigure}{0.45\textwidth}
\includegraphics[width=\textwidth]{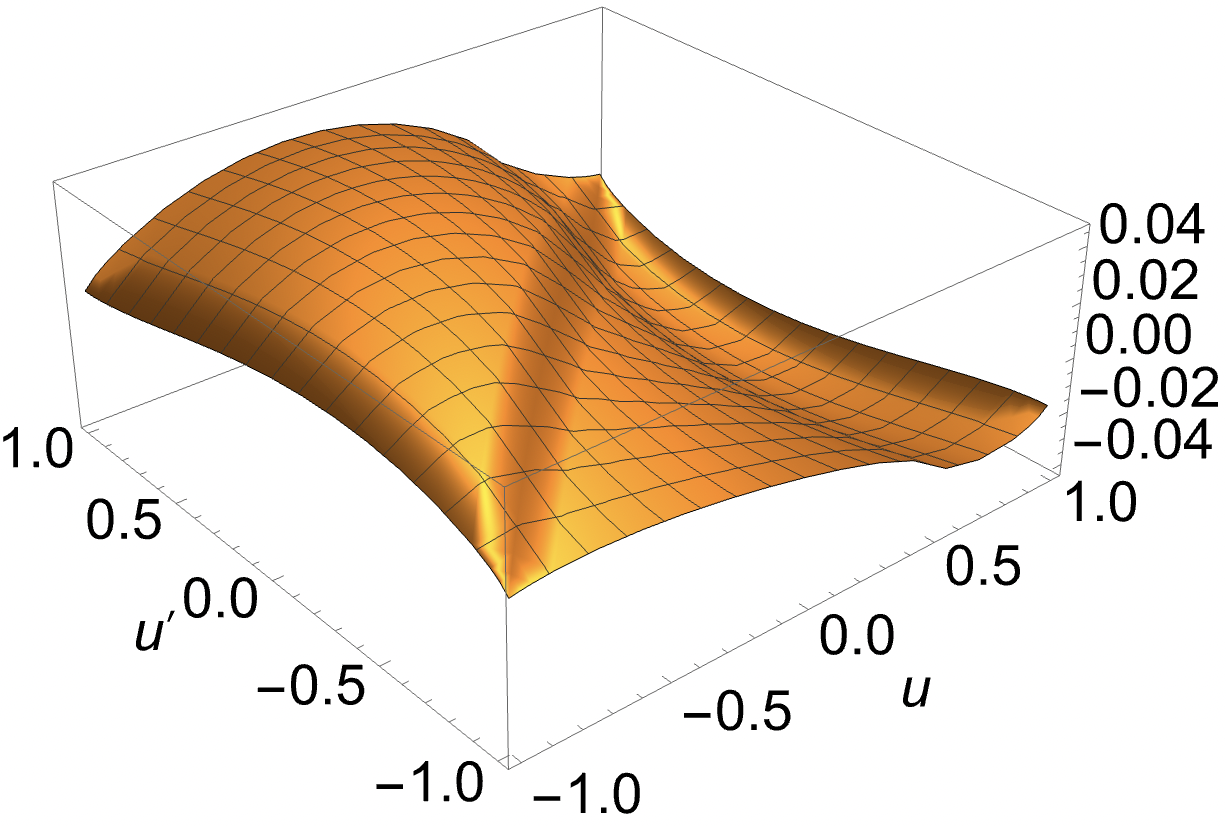}
\caption{Re[$\epsilon^{(2)}_{,u}$]}
\end{subfigure}
\hfill
\begin{subfigure}{0.45\textwidth}
\includegraphics[width=\textwidth]{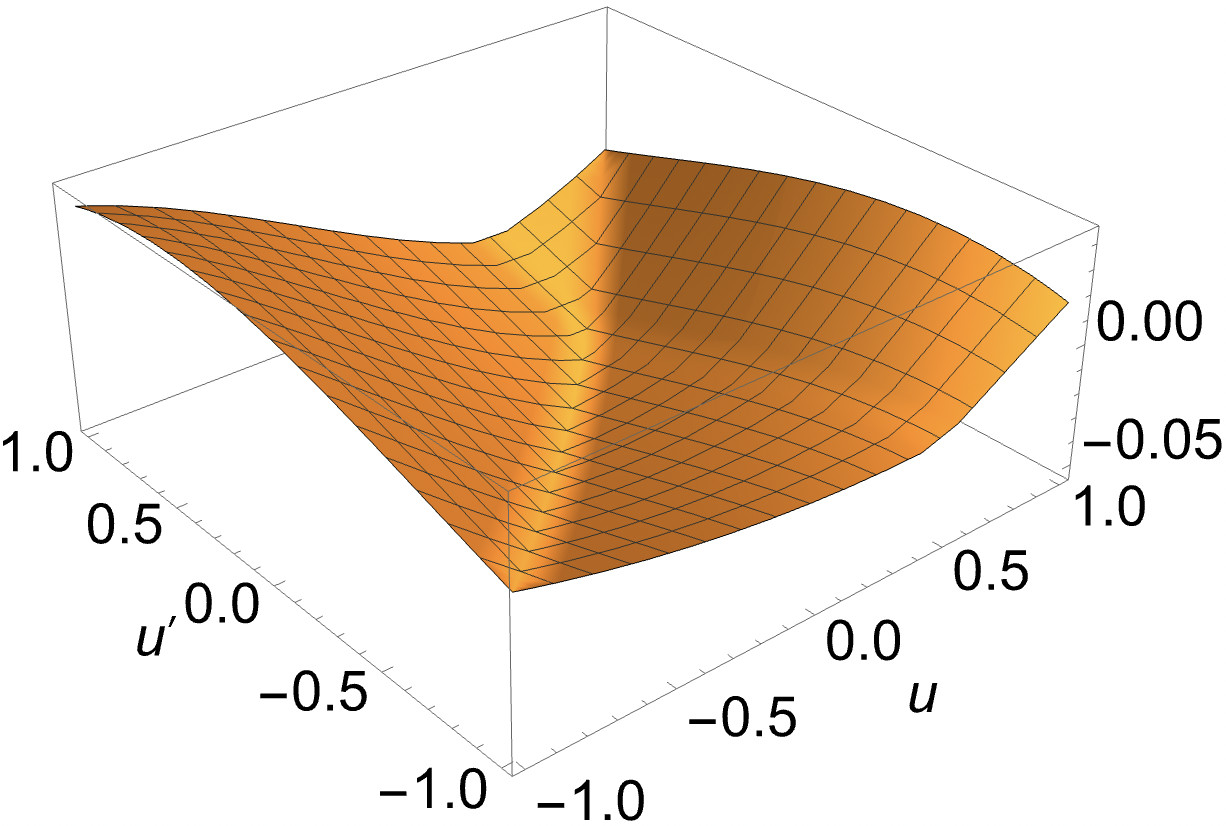}   
\caption{Im[$\epsilon^{(2)}_{,u}$]}
\end{subfigure}

\caption{$\epsilon^{(1)}_{,u}$ and $\epsilon^{(2)}_{,u}$ when $v^\prime = 0.5$. Divergences occur only for Re[$\epsilon^{(1)}_{,u}$] when $u = \pm L$ and $u'\neq\mp L$. $L$ is set to 1.}
\label{fig:epsu}
\end{figure}

 In this subsection we investigate the first derivatives with respect to the coordinates of $\epsilon^{(1)}(u,v;u',v')$ and $\epsilon^{(2)}(u,v;u',v')$. Let us consider $\epsilon_{,u}=\frac{\partial \epsilon} {\partial u}$, the derivative with respect to $u$. The contribution to $\epsilon_{,u}$ from the $n^{th}$ root can be evaluated as
\begin{equation}
\epsilon_{n,u} \equiv \frac{\partial S_{2,n}(k_n,u,v;u^\prime,v^\prime)}{\partial u}
- 
\frac{\partial S_{2,n}( k_n^{(0)},u,v;u^\prime,v^\prime)}{\partial u}.
\label{eq:epsu}
\end{equation}
\noindent
Note that the subtraction and differentiation operations in \eqref{eq:epsu} commute. Hence, we can evaluate $\epsilon_{n,u}$ as either the difference of the derivatives or the derivative of the difference.\footnote{A more subtle issue is taking the derivative of an infinite series.  We assume here that the (partial) derivative $\epsilon_{,u}=\partial_u(\sum_{n=1}^\infty \epsilon_{n})$ equals the infinite series of derivatives $\sum_{n=1}^\infty \epsilon_{n,u}$. Here we merely point out that one way for this equality to hold, is if the sequence of partial sums $\{\sum_{n=1}^M \epsilon_{n,u}\}$  converges \emph{uniformly} on the interval $u\in[-L,L]$ (given the other arguments of $\epsilon$ are held fixed) (see e.g. \cite{rudin}). We will treat other partial derivatives of the series similarly.} Similar to the previous subsection we have
\begin{equation}
\epsilon_{,u} \approx \sum_{j=1}^{\infty} \epsilon_{,u}^{(j)},
\end{equation}
\noindent

where
\begin{equation}
\eqalign{
\epsilon_{,u}^{(j)} &= 
\sum_{n=1}^\infty \epsilon_{n,u}^{(j)} =
\sum_{n=1}^\infty 
\frac{1}{j!} \bigg(\frac{-1}{2k_n^{(0)}L} \bigg)^j
\left.\frac{\partial^j}{\partial k^j}\bigg( \frac{\partial S_{2,n}}{ \partial u} \bigg) \right |_{k = k_n^{(0)}}.  \\
}
\label{eqn:S2Du_n}
\end{equation}
While the expression for $\epsilon^{(1)}$ is quite complicated and difficult to simplify, its first partial derivative can be explicitly evaluated in the nonlocal case to be
\begin{align} \label{eqn:eps1,u}
    &\epsilon^{(1)}_{,u} = \frac{
2iL~\text{arctan}(e^{-\frac{i \pi u}{2L}}) +
(u - u^\prime)~\text{arctanh}[e^{-\frac{i \pi (u - u^\prime)}{2L}}] +
(u - v^\prime)~\text{arctanh}[e^{-\frac{i \pi (u - v^\prime)}{2L}}]
}{8L\pi}.
\end{align}
By considering the individual terms in \eqref{eqn:eps1,u}, we see that the $\text{arctan}$ term has a real logarithmic divergence when $u=\pm L$.  This divergence is cancelled by the logarithmic divergence in either $\text{arctanh}$ terms, when exactly one of $u'$ or $v'$ equals $\mp L$.  $\epsilon^{(2)}_{,u}$ has no divergences, so these are the only divergences observed for $\epsilon_{,u}$, with no imaginary divergences. Similar results hold for $\epsilon_{,v}$, $\epsilon_{,u'}$ and $\epsilon_{,v'}$.  See Figure \ref{fig:epsu} for the example plots of the real and imaginary parts of $\epsilon^{(1)}_{,u}$ and $\epsilon^{(2)}_{,u}$ as a function of $u$ and $u'$ with $v'$ held fixed (as both functions are independent of $v$).  See also Table \ref{tab:nonlocal} for a summary of these results.

\begin{figure}[h]
 \begin{subfigure}{0.32\textwidth}
     \includegraphics[width=\textwidth]{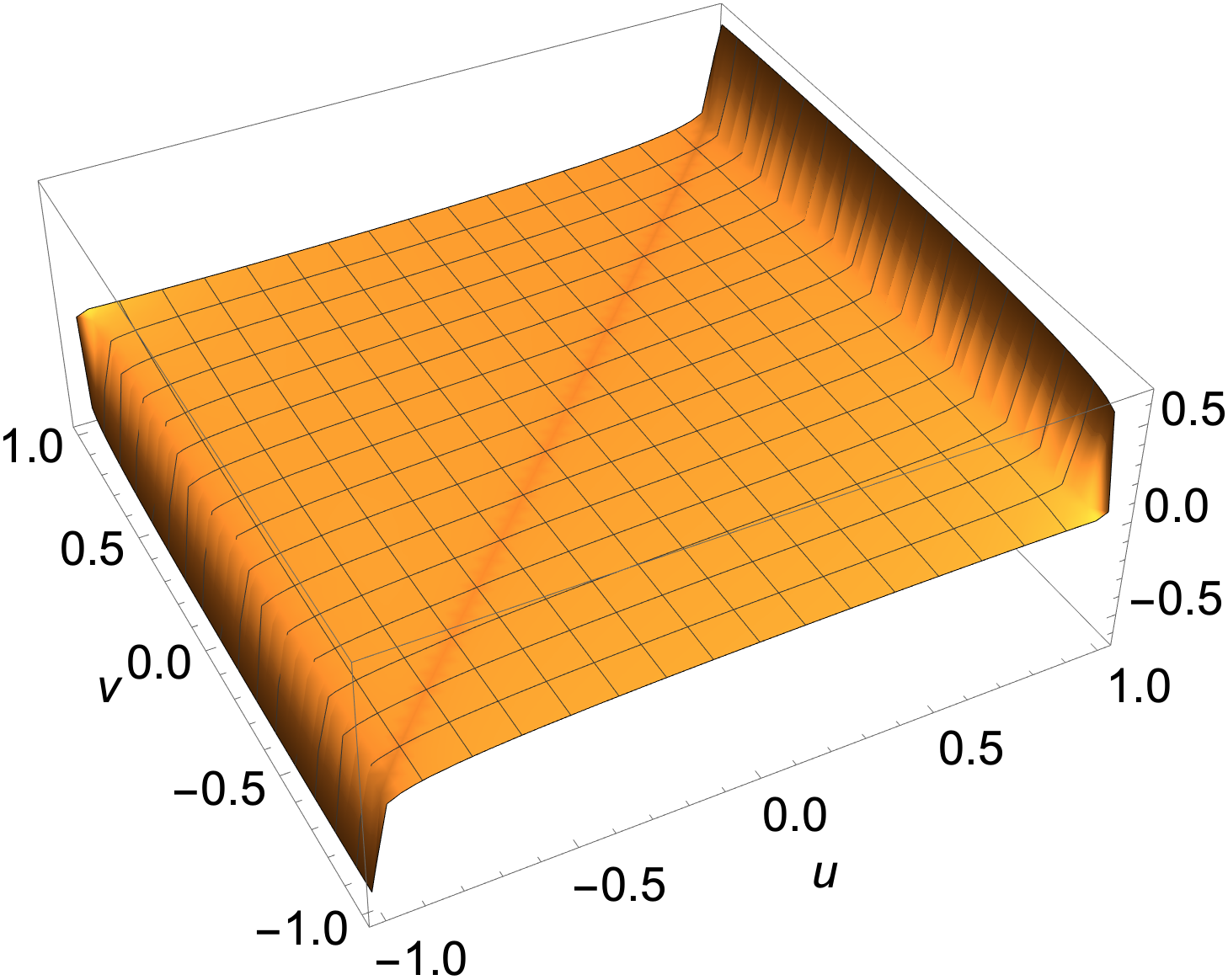}
     \caption{$Re[\epsilon_{,u}^{(1)}]$}
     \label{fig:Du_1_Re}
 \end{subfigure}
 \hfill
 \begin{subfigure}{0.32\textwidth}
     \includegraphics[width=\textwidth]{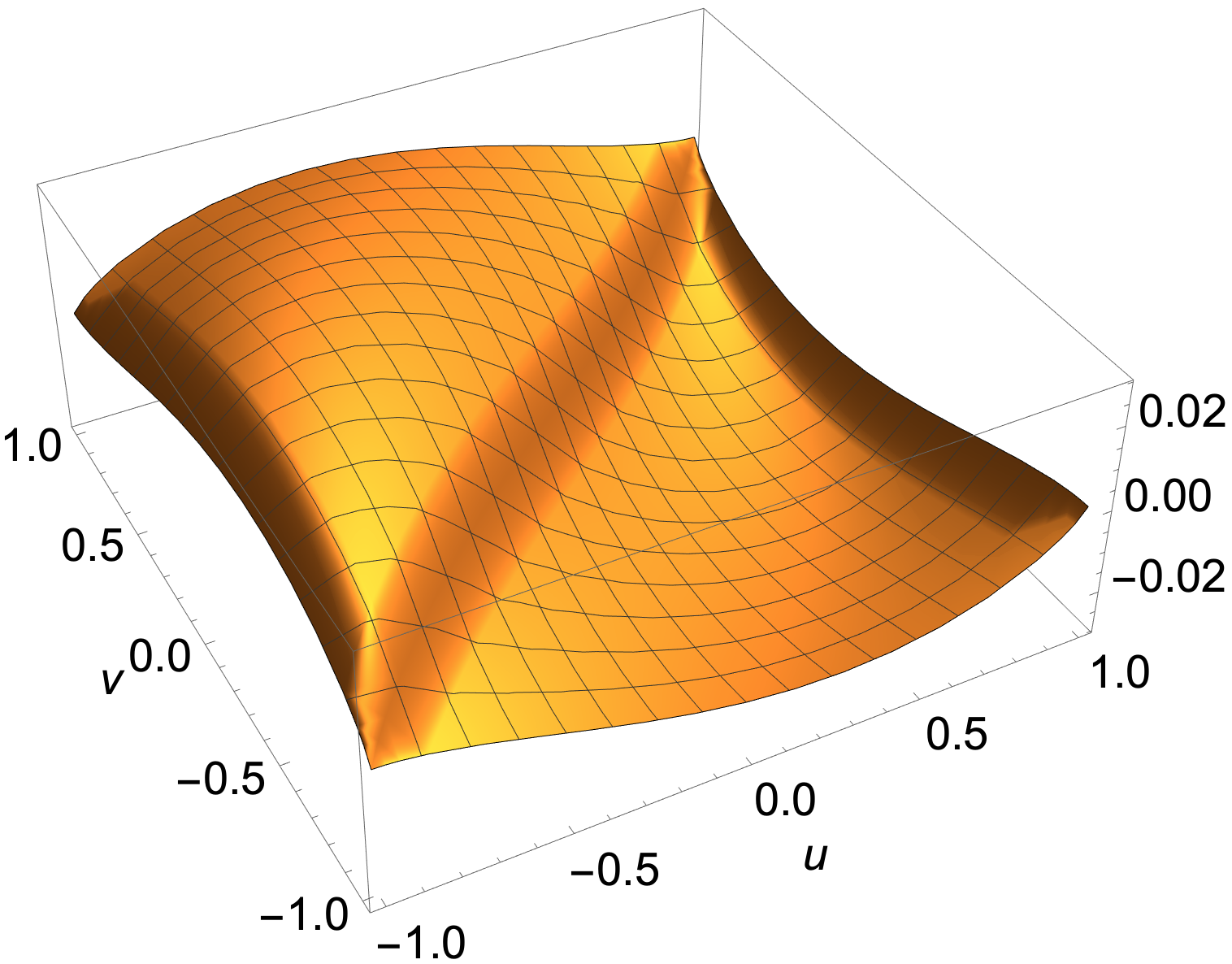}
     \caption{$Re[\epsilon_{,u}^{(2)}]$}
     \label{fig:Du_2_Re}
 \end{subfigure}
  \hfill
 \begin{subfigure}{0.32\textwidth}
     \includegraphics[width=\textwidth]{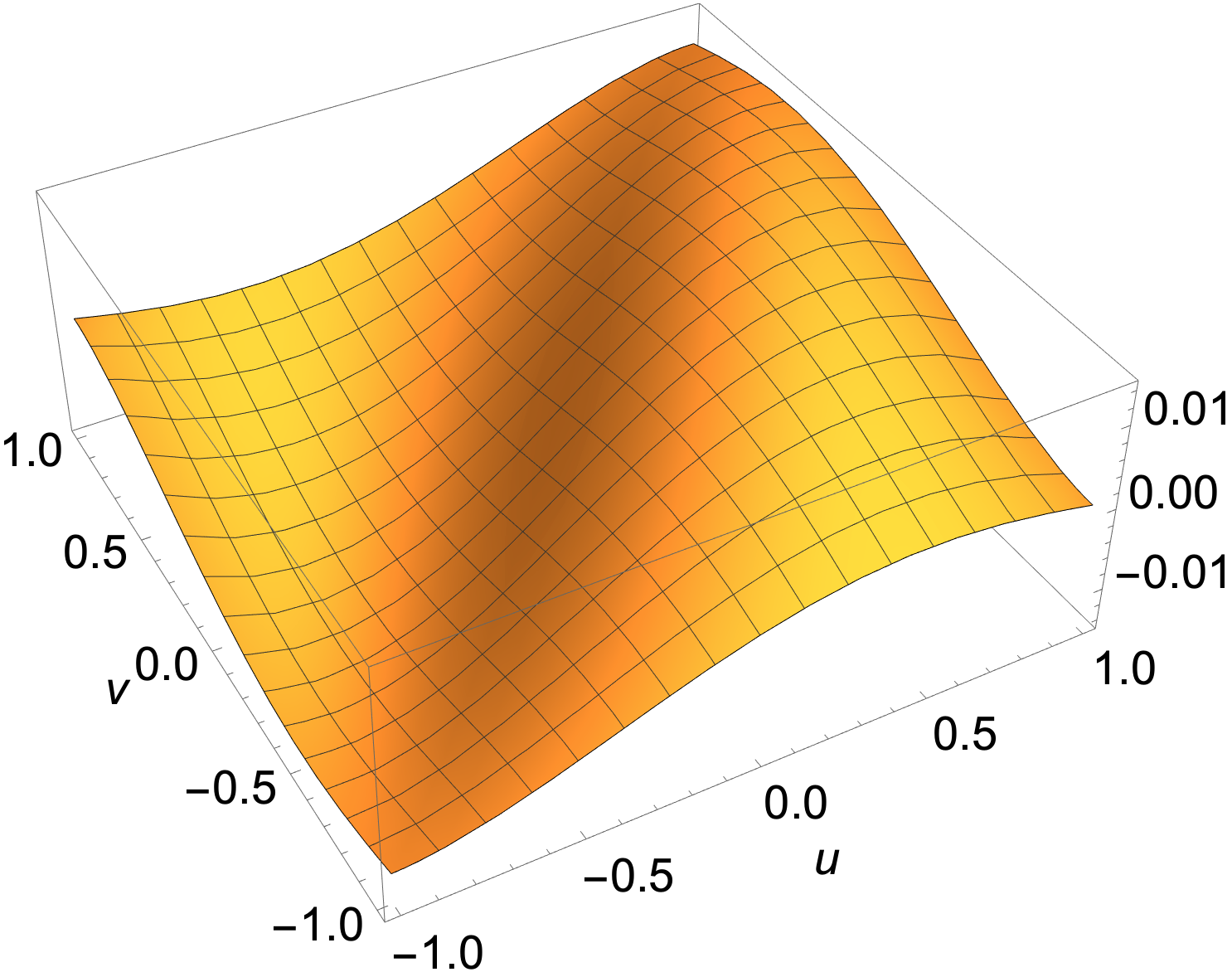}
     \caption{$Re[\epsilon_{,u}^{(3)}]$}
     \label{fig:Du_3_Re}
 \end{subfigure}
 
 \medskip
 \begin{subfigure}{0.32\textwidth}
     \includegraphics[width=\textwidth]{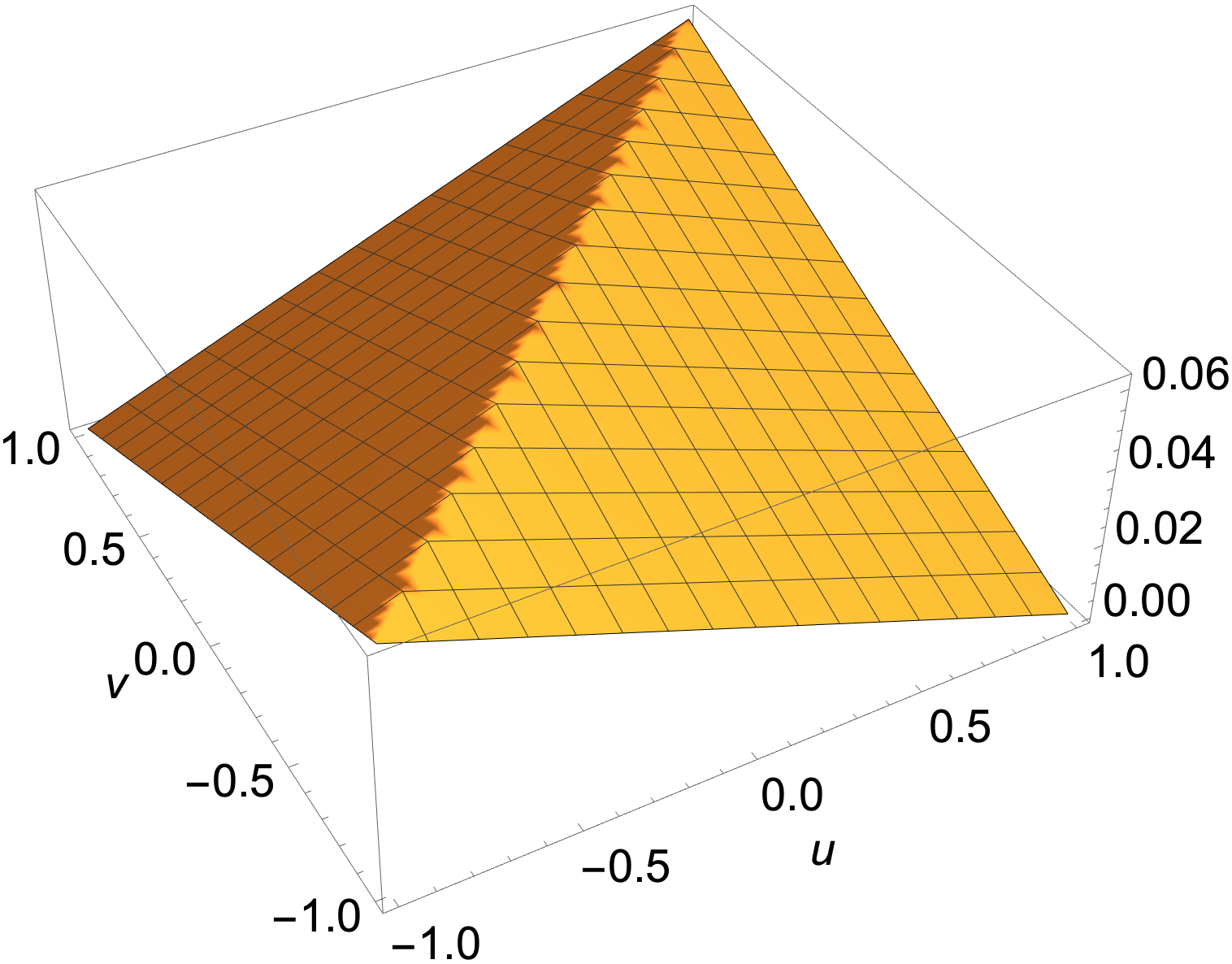}
     \caption{$Im[\epsilon_{,u}^{(1)}]$}
     \label{fig:Du_1_Im}
 \end{subfigure}
 \hfill
 \begin{subfigure}{0.32\textwidth}
     \includegraphics[width=\textwidth]{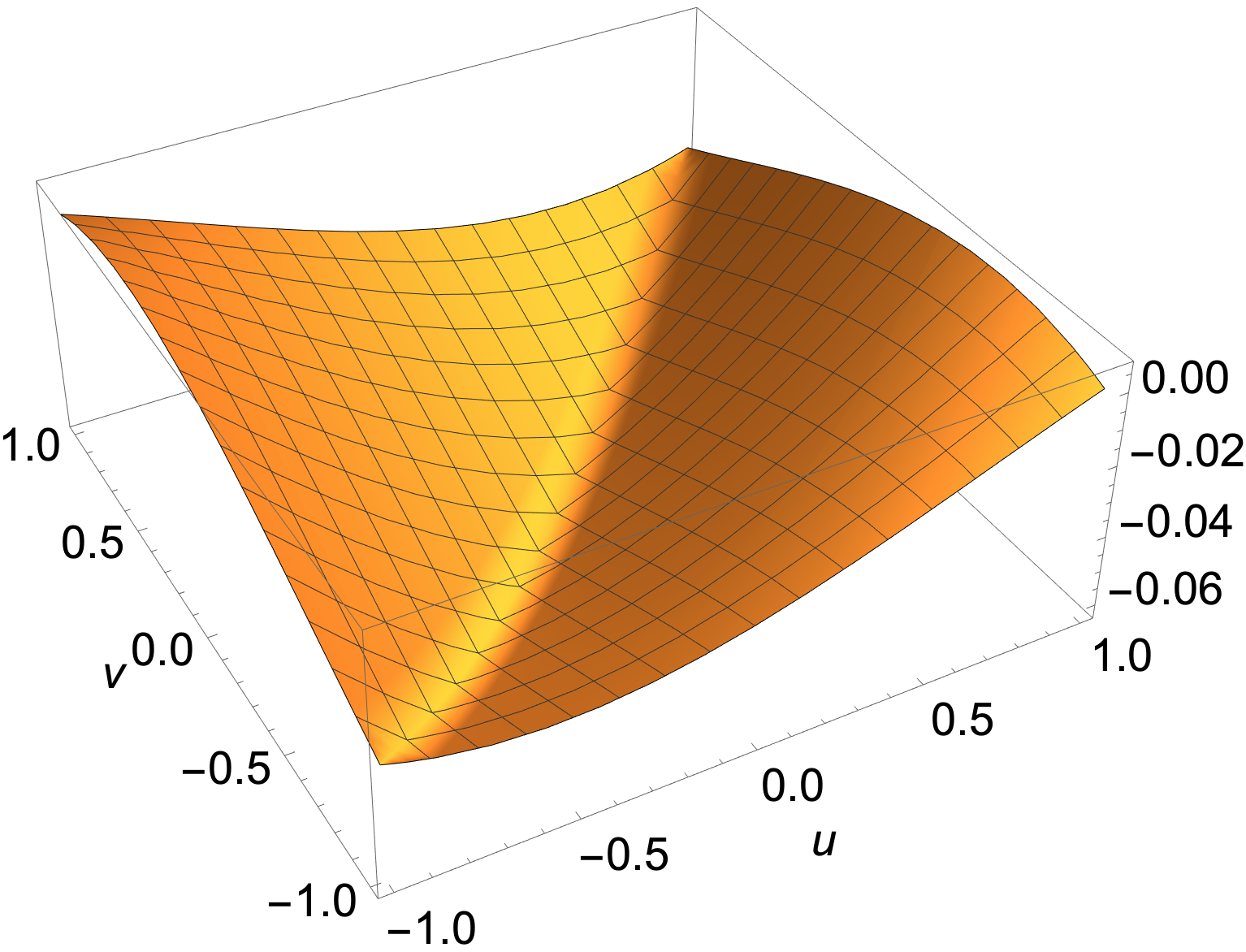}
     \caption{$Im[\epsilon_{,u}^{(2)}]$}
     \label{fig:Du_2_Im}
 \end{subfigure}
 \hfill
 \begin{subfigure}{0.32\textwidth}
     \includegraphics[width=\textwidth]{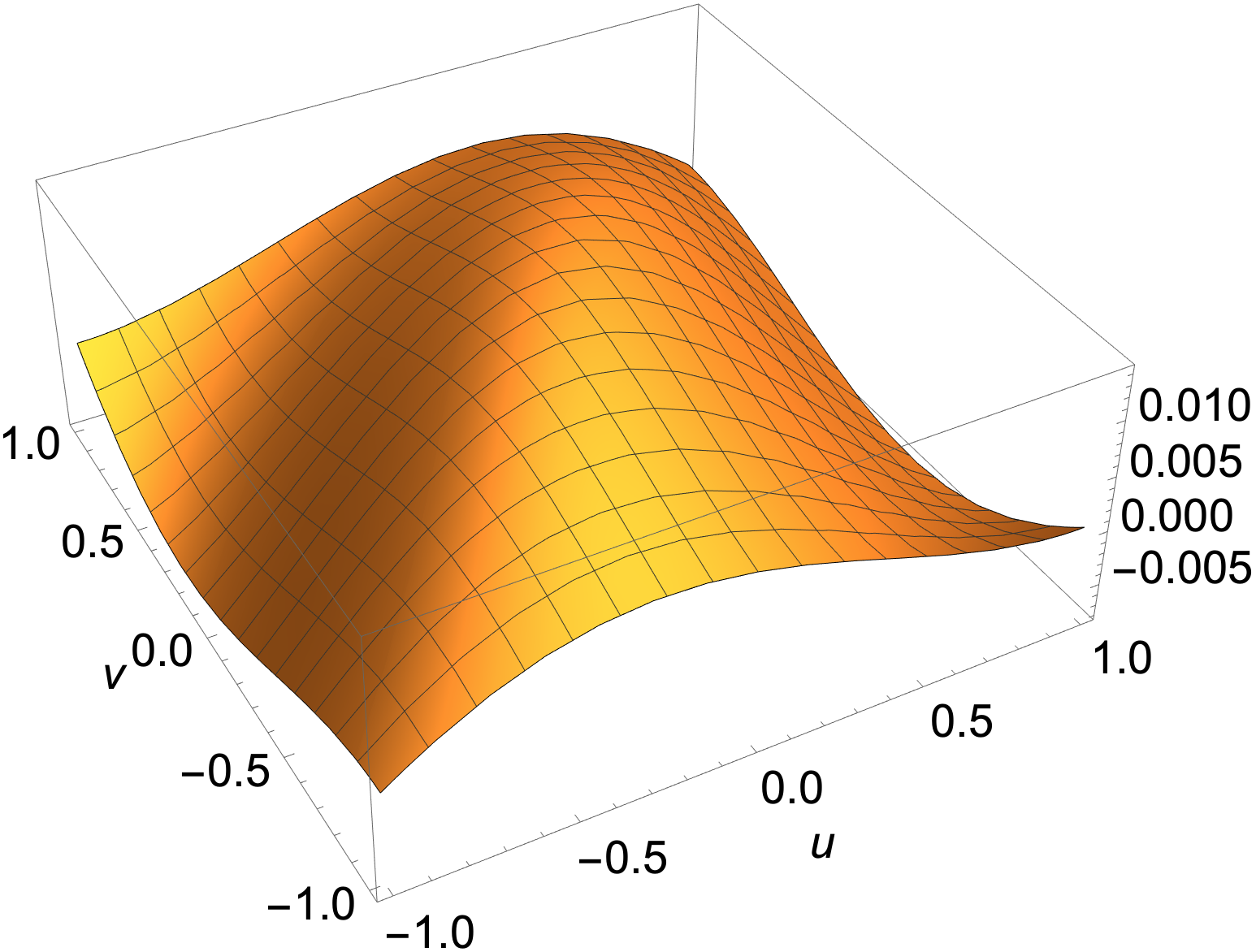}
     \caption{$Im[\epsilon_{,u}^{(3)}]$}
     \label{fig:Du_3_Im}
 \end{subfigure}
 \caption{The first three orders of $\epsilon_{,u}$ in the coincidence limit. A divergence is observed along $u = \pm L$ in the real part of $\epsilon^{(1)}_{,u}$. } 
 \label{fig:Du}
\end{figure}

The coincidence limit is taken after evaluating the derivatives. Setting $L=1$, the values of $\epsilon_{,u}^{(1)}$, $\epsilon_{,u}^{(2)}$ and $\epsilon_{,u}^{(3)}$ are shown in Figure \ref{fig:Du}. Therefore, the only locations of divergence in this limit are along $u = \pm L$ and $v \neq \mp L$\footnote{At $v = \mp L$ the divergence at $u = \pm L$ is cancelled.}, as can be seen in Figure \ref{fig:Du_1_Re}. Due to the symmetry between $u$ and $v$, one would also expect $\epsilon_{,v}$ to diverge along $v = \pm L$ and $u \neq \mp L$. These observations are consistent with the nonlocal divergences mentioned above. The divergences of $\epsilon$ in the coincidence limit are summarized in Table \ref{tab:1}.

We conclude from this that $\epsilon$ is not $C^1$ at the boundary of the diamond, in addition to the zeroth order non-Hadamardness observed in the previous section there.
\subsection{Second Order Derivatives}

There are really only two distinct cases when considering second derivatives of $\epsilon$: The derivative by the same variable twice, and by a unprimed and primed variable.  To be explicit, we will consider $\epsilon_{,uu}$ and $\epsilon_{,u u'}$. In either of the equations below, $u$ can be replaced by $v$ and $u'$ by $v'$ (and vice versa).  This follows from the symmetry of the variables in \eqref{eq:s1sum} and \eqref{eqn:S2}.

\begin{align}
\epsilon^{(1)}_{,uu} = &
\frac{1}{32\,L^2 \, \pi} \bigg\{
 4L \bigg( \textrm{arccoth}\big[ e^{\frac{i \pi(u-u^\prime)}{2L} }
\big] +  \textrm{arccoth}\big[ e^{\frac{i \pi(u-v^\prime)}{2L} }
\big]  \bigg)       
+ \nonumber\\
& \pi(u^\prime - u)\,\textrm{cosec}\bigg[\frac{\pi(u - u^\prime)}{2L} \bigg] + \pi(v^\prime - u)\,\textrm{cosec}\bigg[\frac{\pi(u - v^\prime)}{2L} \bigg] + 2 L\pi\, \textrm{sec}\bigg( \frac{\pi u}{2L} \bigg)
\bigg\}.
\label{eqn:epsuu1}
\end{align}

\begin{figure}[h]
\begin{subfigure}{0.45\textwidth}
\includegraphics[width=\textwidth]{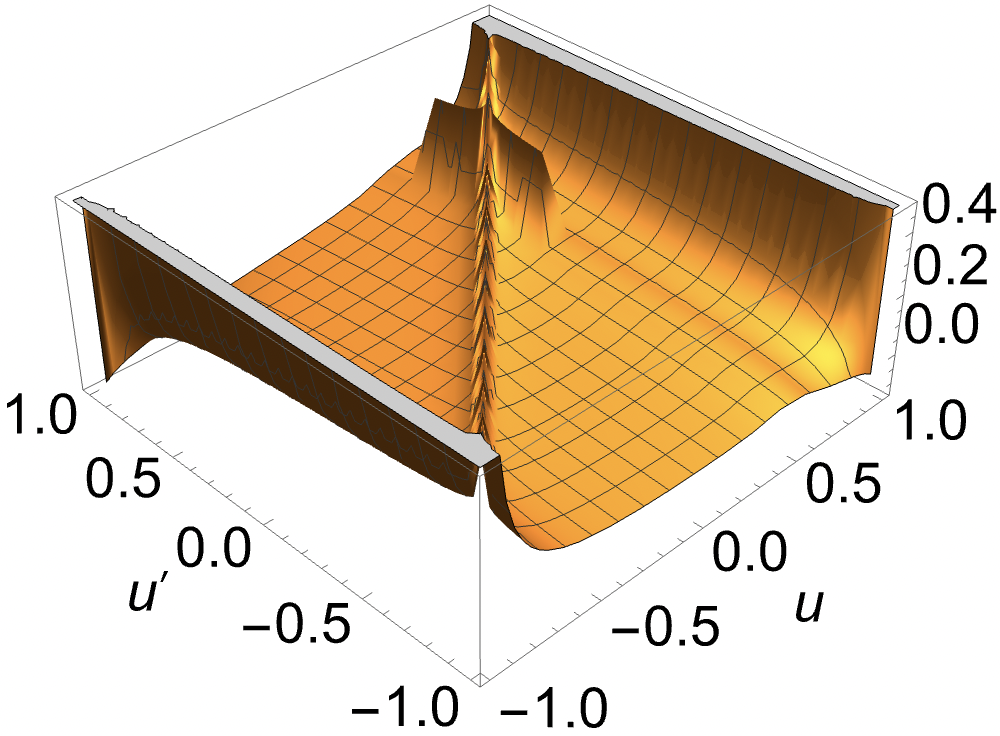}
\caption{Re[$\epsilon^{(1)}_{,uu}$]}
\end{subfigure}
\hfill
\begin{subfigure}{0.45\textwidth}
\includegraphics[width=\textwidth]{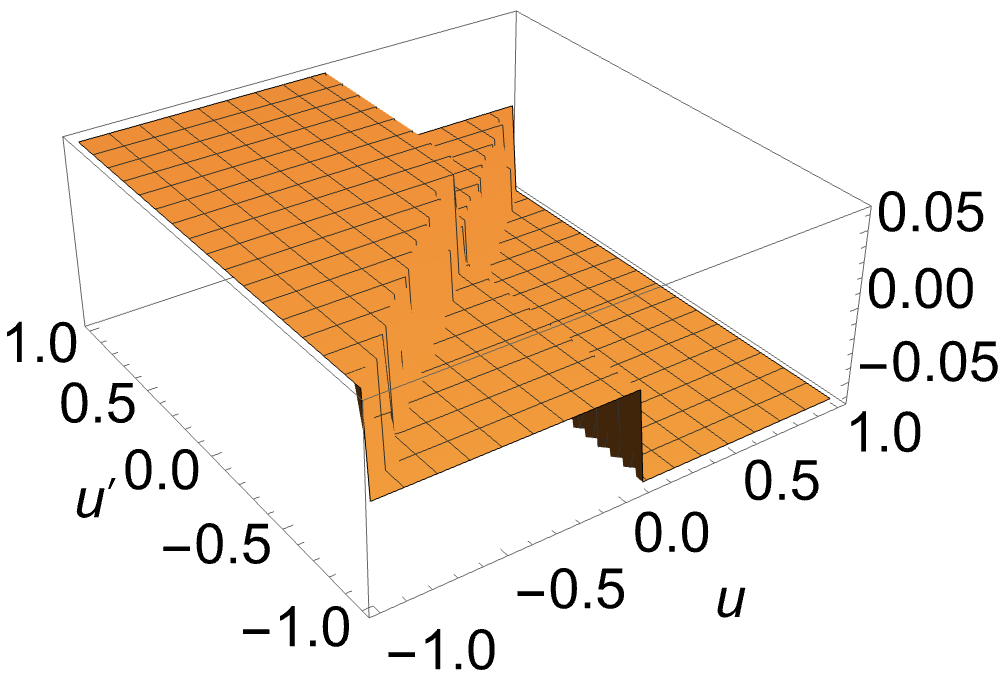}   
\caption{Im[$\epsilon^{(1)}_{,uu}$]}
\end{subfigure}

\medskip

\begin{subfigure}{0.45\textwidth}
\includegraphics[width=\textwidth]{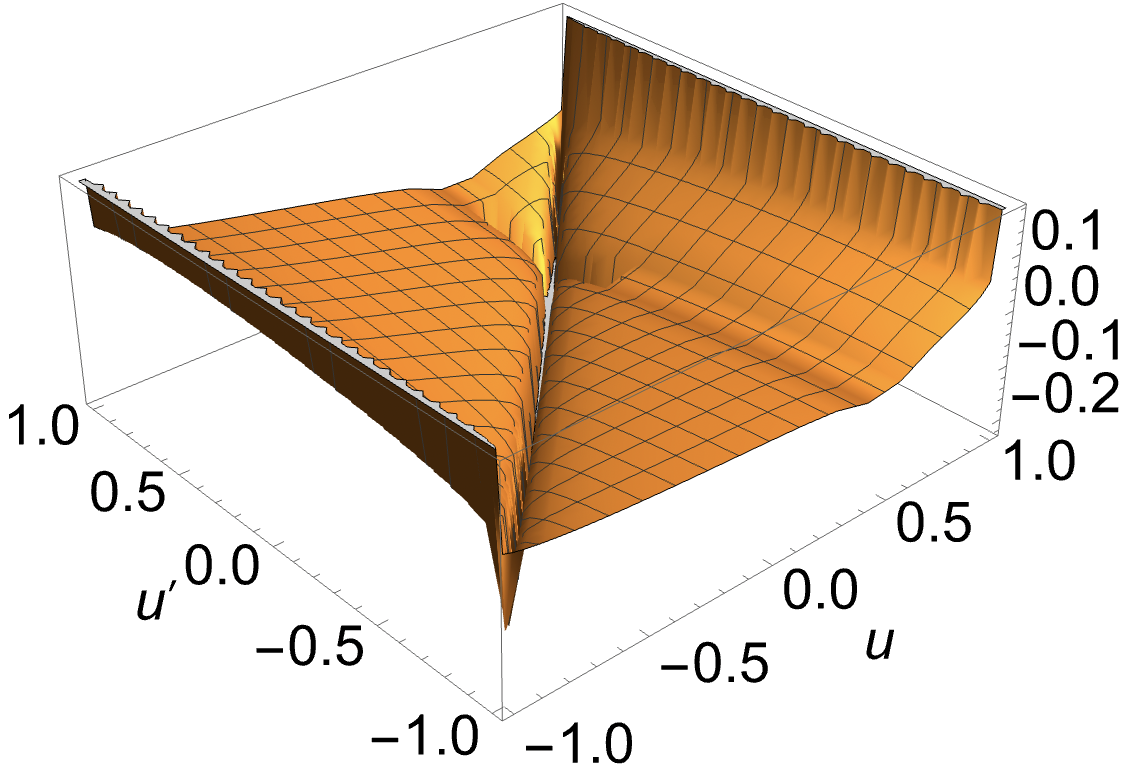}
\caption{Re[$\epsilon^{(2)}_{,uu}$]}
\end{subfigure}
\hfill
\begin{subfigure}{0.45\textwidth}
\includegraphics[width=\textwidth]{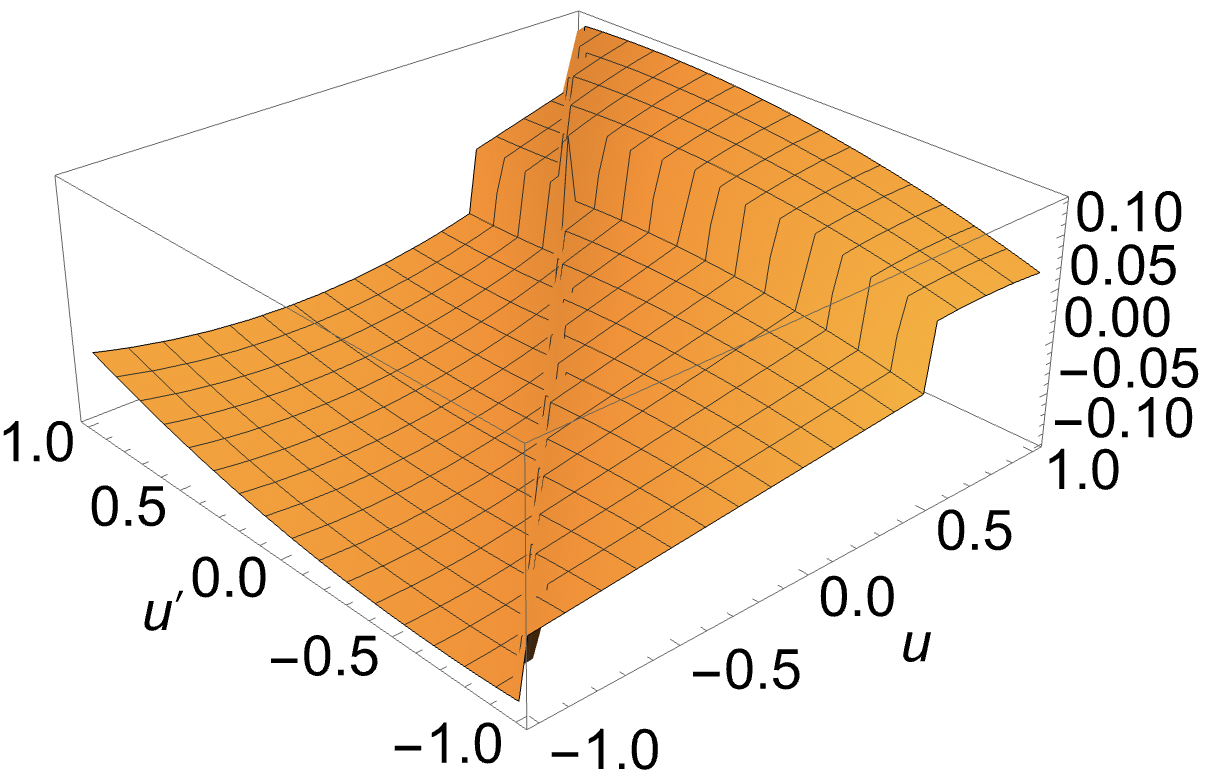}   
\caption{Im[$\epsilon^{(2)}_{,uu}$]}
\end{subfigure}

\caption{$\epsilon^{(1)}_{,uu}$ and $\epsilon^{(2)}_{,uu}$ when $v^\prime = 0.5 L$. $L$ is set to be 1 in the plots. Both real parts of $\epsilon^{(1)}_{,uu}$ and $\epsilon^{(2)}_{,uu}$ diverge at $u=\pm L$ (the cancellation when $(u,u^\prime,v^\prime) = (\pm L, \mp L, \pm L)$ and $(\pm L, \pm L, \mp L)$ are not visible here as $v'$ is set to $0.5 L$).  The divergences along $u = u^\prime$ cancel out after taking the sum of $\epsilon^{(1)}_{,uu}$ and $\epsilon^{(2)}_{,uu}$.}
\label{fig:epsuu}
\end{figure}

Both $\epsilon^{(1)}_{,uu}$ and $\epsilon^{(2)}_{,uu}$ diverge when $u = \pm L$ and $u = u^\prime, v^\prime$. The $u = u^\prime, v^\prime$ divergences in $\epsilon^{(1)}_{,uu}$ come from the two arccoth terms in \eqref{eqn:epsuu1}.  This logarithmic divergence is exactly cancelled by the divergence in $\epsilon^{(2)}_{,uu}$.  The $u = \pm L$ divergence in $\epsilon^{(1)}_{,uu}$ is inverse-linear and originates from the sec term, and is not cancelled by $\epsilon^{(2)}_{,uu}$ as both functions diverge with the same sign.  The divergence in the sec term is cancelled when $u'$ or $v'=\mp L$ from the cosec terms. However, this leads to a logarithmic divergence coming from the corresponding arccoth term unless the remaining primed variable has the opposite sign, in which case the logarithmic divergences from the two arccoth terms also cancel.  To summarize, $\epsilon^{(1)}_{,uu}$ has a divergence at $u = \pm L$ except when $(u,u^\prime,v^\prime) = (\pm L, \mp L, \pm L)$ or $(\pm L, \pm L, \mp L)$.  $\epsilon^{(2)}_{,uu}$ is also finite at these points. See Figure \ref{fig:epsuu} for plots of the real and imaginary parts of  $\epsilon^{(1)}_{,uu}$ and  $\epsilon^{(2)}_{,uu}$ vs $u$ and $u'$ for a fixed $v'$.  Note that there are no imaginary divergences.

\begin{figure}[h]
\begin{subfigure}{0.45\textwidth}
\includegraphics[width=\textwidth]{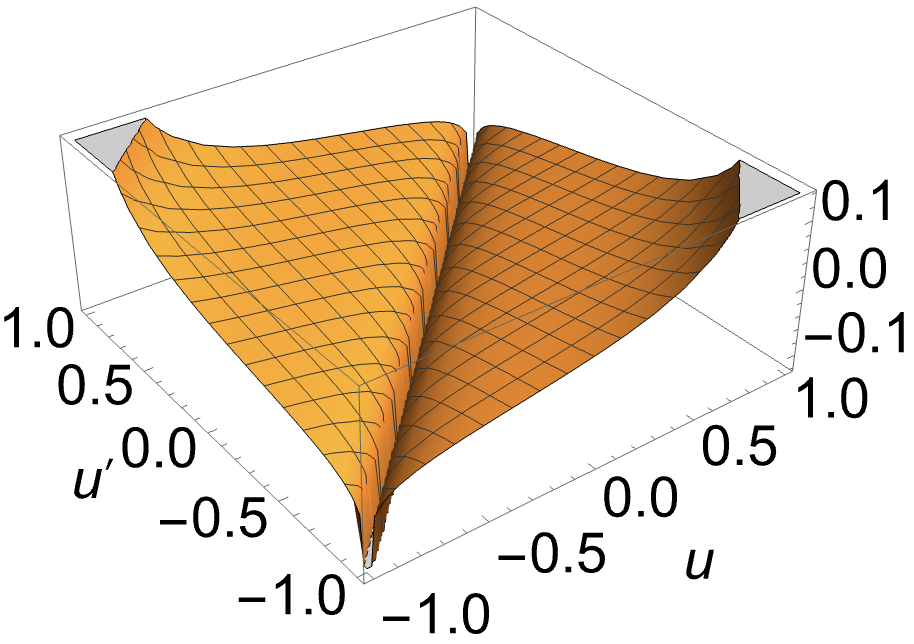}
\caption{Re[$\epsilon^{(1)}_{,uu^\prime}$]}
\end{subfigure}
\hfill
\begin{subfigure}{0.45\textwidth}
\includegraphics[width=\textwidth]{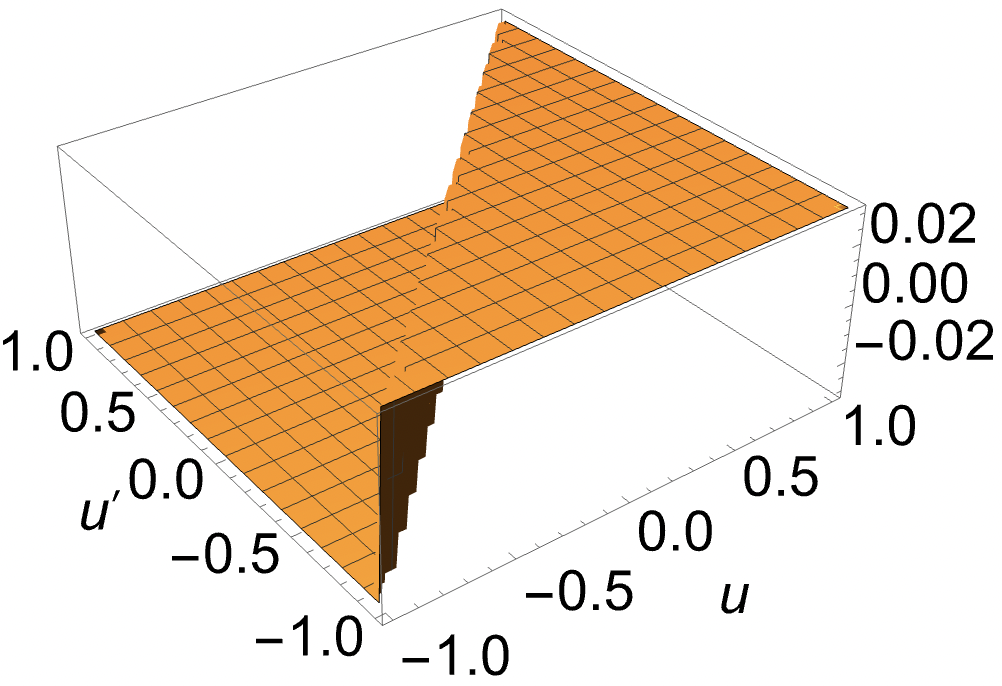}   
\caption{Im[$\epsilon^{(1)}_{,uu^\prime}$]}
\end{subfigure}

\medskip

\begin{subfigure}{0.45\textwidth}
\includegraphics[width=\textwidth]{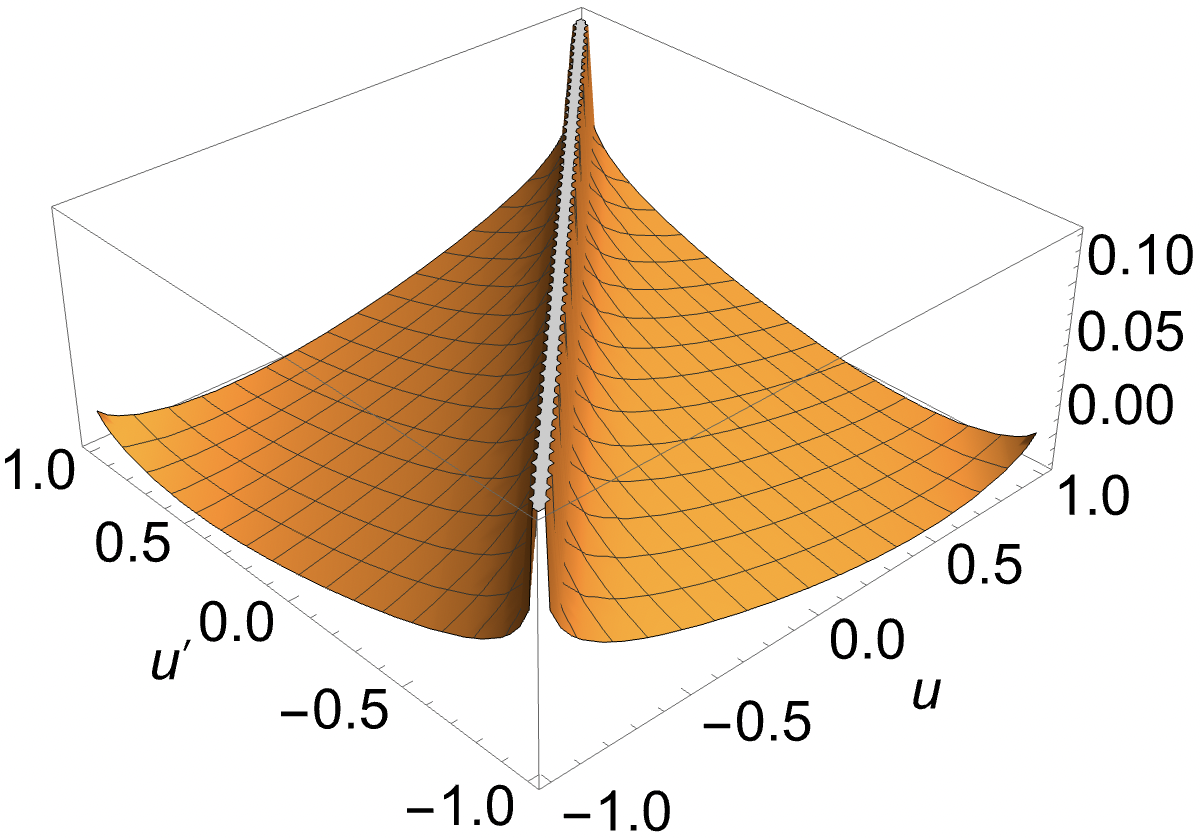}
\caption{Re[$\epsilon^{(2)}_{,uu^\prime}$]}
\end{subfigure}
\hfill
\begin{subfigure}{0.45\textwidth}
\includegraphics[width=\textwidth]{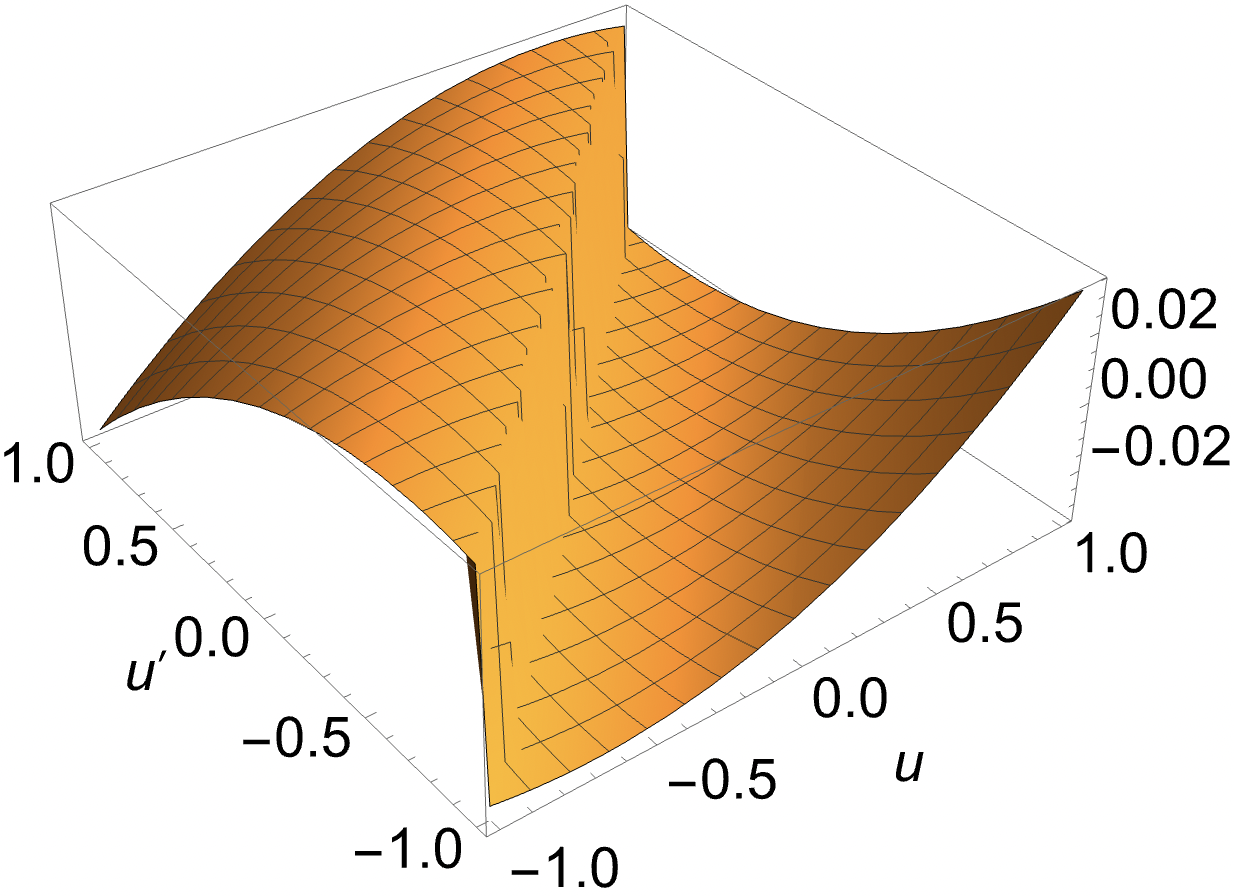}   
\caption{Im[$\epsilon^{(2)}_{,uu^\prime}$]}
\end{subfigure}

\caption{$\epsilon^{(1)}_{,uu^\prime}$ and $\epsilon^{(2)}_{,uu^\prime}$ (both functions only depend on $u$ and $u^\prime$). The real part of $\epsilon^{(1)}_{,uu^\prime}$ diverges when $(u,u^\prime) = (\pm L, \mp L)$. The divergences along $u = u^\prime$ cancel out after taking the sum of $\epsilon^{(1)}_{,uu^\prime}$ and $\epsilon^{(2)}_{,uu^\prime}$. $L$ is set to be 1 in the plots.}
\label{fig:epsua}
\end{figure}

We also have a closed form expression for $\epsilon^{(1)}_{,uu^\prime}$
\begin{align}
&\epsilon^{(1)}_{,uu^\prime} = \frac{-4L\,\textrm{arccoth}
\big[ e^\frac{i \pi (u - u^\prime)}{2L} \big] + \pi(u - u^\prime)\,\textrm{cosec}\big[ \frac{\pi(u-u^\prime)}{2L} \big]
}{32\,L^2 \, \pi}.
\label{eqn:epsua}
\end{align}
In the above equation we find that there are divergences when $u=u'$ and when $u=\pm L$ and $u'=\mp L$.  As before the $u=u'$ divergence cancels between $\epsilon^{(1)}_{,uu^\prime}$ and $\epsilon^{(2)}_{,uu^\prime}$.  However, the divergences at $u=\pm L$ and $u'=\mp L$ do not cancel.  See Figure \ref{fig:epsua} for plots of the real and imaginary parts of $\epsilon^{(1)}_{,uu^\prime}$ and $\epsilon^{(2)}_{,uu^\prime}$, and Table \ref{tab:nonlocal} for a summary of all nonlocal divergences of the derivatives of $\epsilon$.

The above analysis of course translates to the coincidence limit. The results are summarized in Table \ref{tab:1}. As with the first derivative, the divergences in this limit all occur on the boundary of the diamond (in the nonlocal case at least one of the points  $(u, v)$, $(u', v')$ must lie on the boundary of the diamond).

\begin{tabularx}{\textwidth}{|X|X|X|X|}
\hline
          & Dependencies &  Re & Im \\
 \hline
 $\epsilon^{(1)}_{,u}$ +  $\epsilon^{(2)}_{,u}$ &  $u,u^\prime, v^\prime$ & $u = \pm L$  except when one of $u^\prime$ or $v^\prime$ = $\mp L$ & /  \\
 \hline
 $\epsilon^{(1)}_{,v}$ +  $\epsilon^{(2)}_{,v}$ &  $v,u^\prime, v^\prime$ 
 & $v = \pm L$  except when one of $u^\prime$ or $v^\prime$ = $\mp L$ & /  \\
 \hline
 $\epsilon^{(1)}_{,uu}$ +  $\epsilon^{(2)}_{,uu}$ &  $u,u^\prime, v^\prime$  & $u = \pm L$ except when $(u,u^\prime,v^\prime) = (\pm L, \mp L, \pm L)$ or  $(\pm L, \pm L, \mp L)$  & / \\
 \hline
 $\epsilon^{(1)}_{,vv}$ +  $\epsilon^{(2)}_{,vv}$ &  $v,u^\prime, v^\prime$   & $v = \pm L$ except when $(v,u^\prime,v^\prime) = (\pm L, \mp L, \pm L)$ or  $(\pm L, \pm L, \mp L)$ & / \\
 \hline
 $\epsilon^{(1)}_{,uu^\prime}$ +  $\epsilon^{(2)}_{,uu^\prime}$ &  $u,u^\prime $  & $(u,u^\prime) = (\pm L, \mp L)$  & / \\
 \hline
 $\epsilon^{(1)}_{,vv^\prime}$ +  $\epsilon^{(2)}_{,vv^\prime}$  &  $v,v^\prime $& $(v,v^\prime) = (\pm L, \mp L)$  & /  \\
 \hline
 $\epsilon^{(1)}_{,uv^\prime}$ +  $\epsilon^{(2)}_{,uv^\prime}$  &  $u,v^\prime$ & $(u,v^\prime) = (\pm L, \mp L)$  & / \\
 \hline
 $\epsilon^{(1)}_{,vu^\prime}$ +  $\epsilon^{(2)}_{,vu^\prime}$ &  $v,u^\prime$  & $(v,u^\prime) = (\pm L, \mp L)$  & / \\
\hline
 \caption{The locations of divergences of the first and second order derivatives of $\epsilon$, as well as the the variables they depend on, when expanded up to second order in \eqref{eqn:epsilon_approx}.}\label{tab:nonlocal}
 \end{tabularx}

\begin{tabularx}{\textwidth}{|X|X|X|}
\hline
          & Re & Im \\
 \hline
 $\epsilon_{,u}$  & $u = \pm L$ \& $v \neq \mp L$  & /  \\
 \hline
 $\epsilon_{,v}$  & $v = \pm L$  \& $u \neq \mp L$ & /  \\
 \hline
 $\epsilon_{,uu}$  &$u = \pm L$ \& $v \neq \mp L$  &  /  \\
 \hline
 $\epsilon_{,vv}$  &  $v = \pm L$  \& $u \neq \mp L$ &  /\\
 \hline
 $\epsilon_{,uu^\prime}$  & /  & /  \\
 \hline
 $\epsilon_{,vv^\prime}$  & /  & /  \\
 \hline
 $\epsilon_{,uv^\prime}$  & ($\pm L, \mp L$)  & / \\
 \hline
 $\epsilon_{,vu^\prime}$  & ($\pm L, \mp L$)  &  / \\
\hline
 \caption{The locations of divergence of the first and second derivatives of $\epsilon$ with respect to the lightcone coordinates $(u,v)$ in the coincidence limit.}\label{tab:1}
\end{tabularx}

\section{Entanglement Entropy with the Softened SJ State}
\label{sec:Entropy}

Sorkin has suggested that the non-Hadamard nature of the SJ vacuum stems from the fact that in a finite time interval (such as in a diamond), positive and negative frequencies are not
orthogonal in the $L^2$ inner product \cite{Sorkin:2017fcp}. Namely,  $\langle\phi|\bar{\phi}\rangle\neq 0$, as it would be if time extended from $-\infty$ to $\infty$. Based on this intuition, a \emph{softening of the boundary} was suggested in order to make the SJ state Hadamard: one  modifies the $L^2$ norm used in defining the inner product of the eigenfunctions of $i\Delta$ by introducing a different integration
measure. To this end, we can modify the SJ eigenvalue equation for $i\Delta$ from 
\begin{equation}
 \int i\Delta(x,x') T_k(x') dV_{x'}=\lambda_k T_k(x)
\end{equation}

to

\begin{equation}
 \int i\Delta(x,x') \tilde{T}_k(x') \rho(x') dV_{x'}=\tilde{\lambda}_k \tilde{T}_k(x),
 \label{eq:softcontinuum}
\end{equation}
 where $\rho(x')$ is chosen to go smoothly to zero at
the boundary. As there is no preferred specific expression for $\rho(x')$, this softening leads to many different modified SJ states, one for each choice of $\rho(x')$. 

\subsection{Softened SJ State in a Causal Set Diamond}
In the discrete causal set which we will work with, \eqref{eq:softcontinuum} becomes the sum

\begin{equation}
  \sum_{x'} \rho(x') i \Delta(x,x^\prime)T_q^+(x^\prime)=\tsup[1]{\lambda}_q \tsup[1]{T}_q^+(x) .
\end{equation}
We can then define the softened SJ Wightman function, $\tsup[1]{W}_{SJ}$, as
\begin{equation}
\tsup[1]{W}_{SJ}(x,x^\prime) = \sum_{q} \tsup[1]{\lambda}_q \tsup[1]{T}_q^+(x) \tsup[1]{T}_q^+(x^\prime)^*.
\end{equation}

This procedure is known to make the SJ state Hadamard in the continuum \cite{Wingham:2018pxx}. 

Of course, even the softened SJ state in the causal set is not strictly Hadamard in the usual sense of the condition, as the causal set always modifies the short distance (near the discreteness scale) behavior away from continuum Minkowski spacetime. However, in the same spirit of the  desire to minimally deform from the Minkowski state, we can view the softened states in the causal set to be minimal deformations of the Minkowski state restricted to a Minkowski spacetime sprinkling.

\begin{figure}[h]
\centering
\includegraphics[width=10cm]{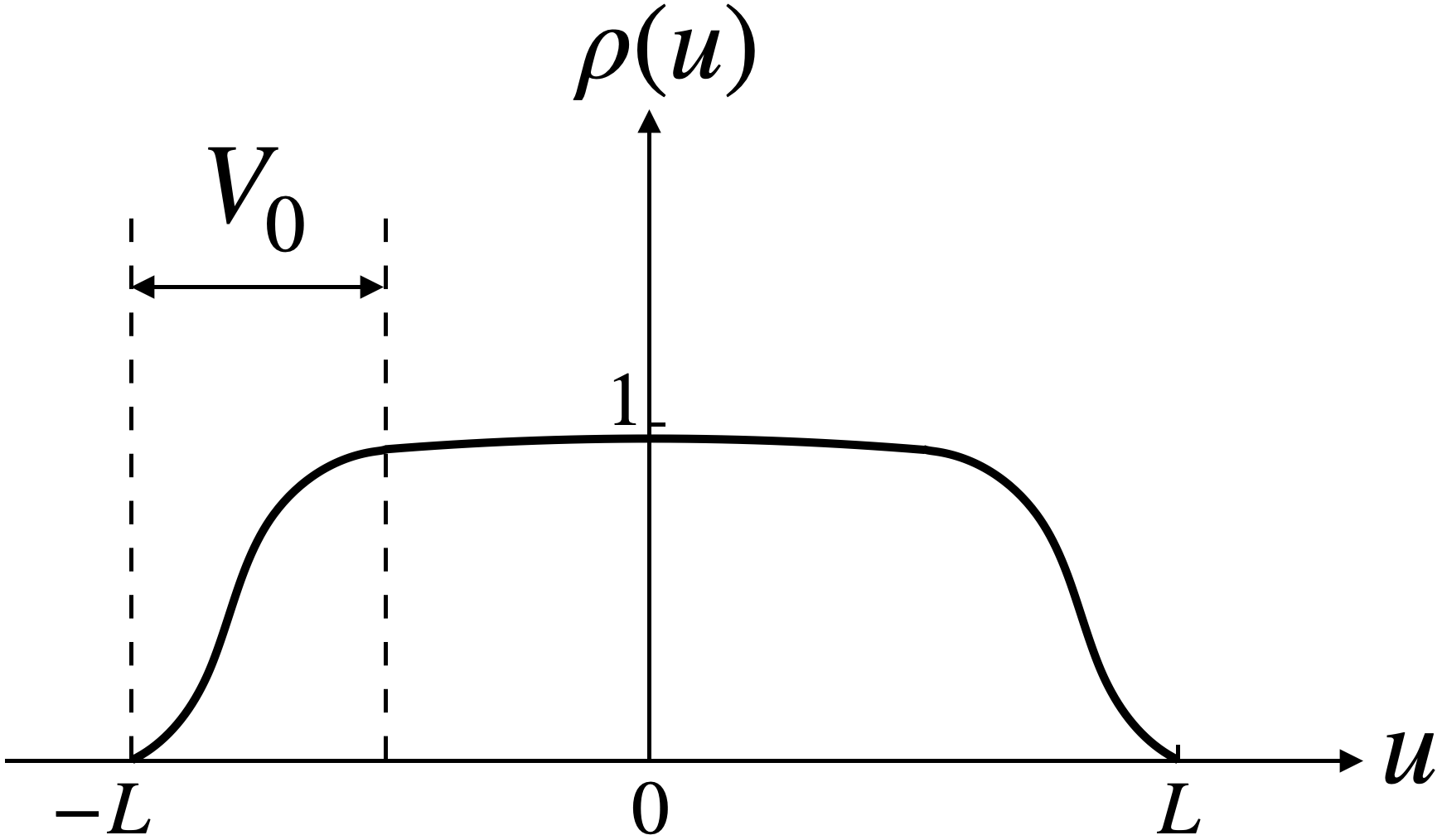}
\caption{ The softening function $\rho$ as a function of the lightcone coordinate $u$. $V_0$ sets the scale of the softening. The value of $\rho$ in the centre of the causal diamond should remain close to 1 such that there is  minimal change to the Minkowski-like behavior there.}
\label{fig:rho}
\end{figure}

We will use a particular convenient functional form for $\rho(x')$ suggested to us by Sorkin \cite{private_email}. Using the index $i$ to label the causal set elements, we define $\rho_i$ to be
\begin{equation}
\rho_i= 1- e^{-\frac{r_i s_i}{r_i+s_i+1}},
\end{equation}
 where
\begin{equation} \label{softenening_defs}
r_i = \frac{V^+_i}{V_0}, \quad s_i = \frac{V^-_i}{V_0} ,
\end{equation}
and where $V^\pm_i=|\mathcal{J}^\pm (i)|$ or the future and past volumes of the element $i$. $V_0$ sets the scale of the softening (i.e.  how close one
gets to the boundary before $\rho$ decays). $V_0\rightarrow 0$ is the limit to the regular SJ prescription with a hard boundary. See for example Figure \ref{fig:rho}.

\begin{figure}[h]
\centering
\includegraphics[width=10cm]{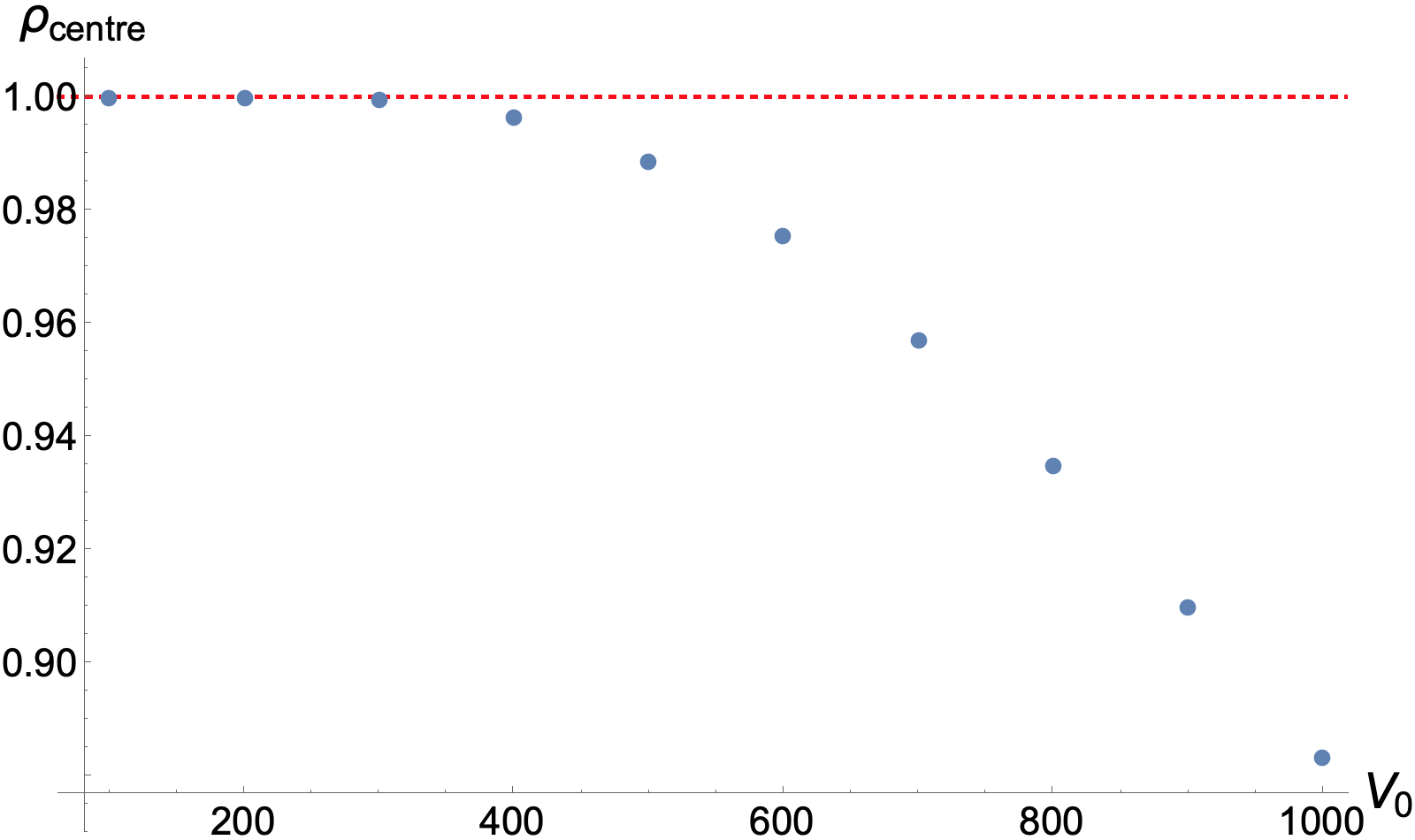}
\caption{The average value of $\rho$ in the centre of the causal set diamond $|t|\leq0.28 L$, $|x|\leq0.28L$) versus $V_0$. }
\label{fig:VRho}
\end{figure}

\begin{figure}[h]
\centering
\includegraphics[width=13cm]{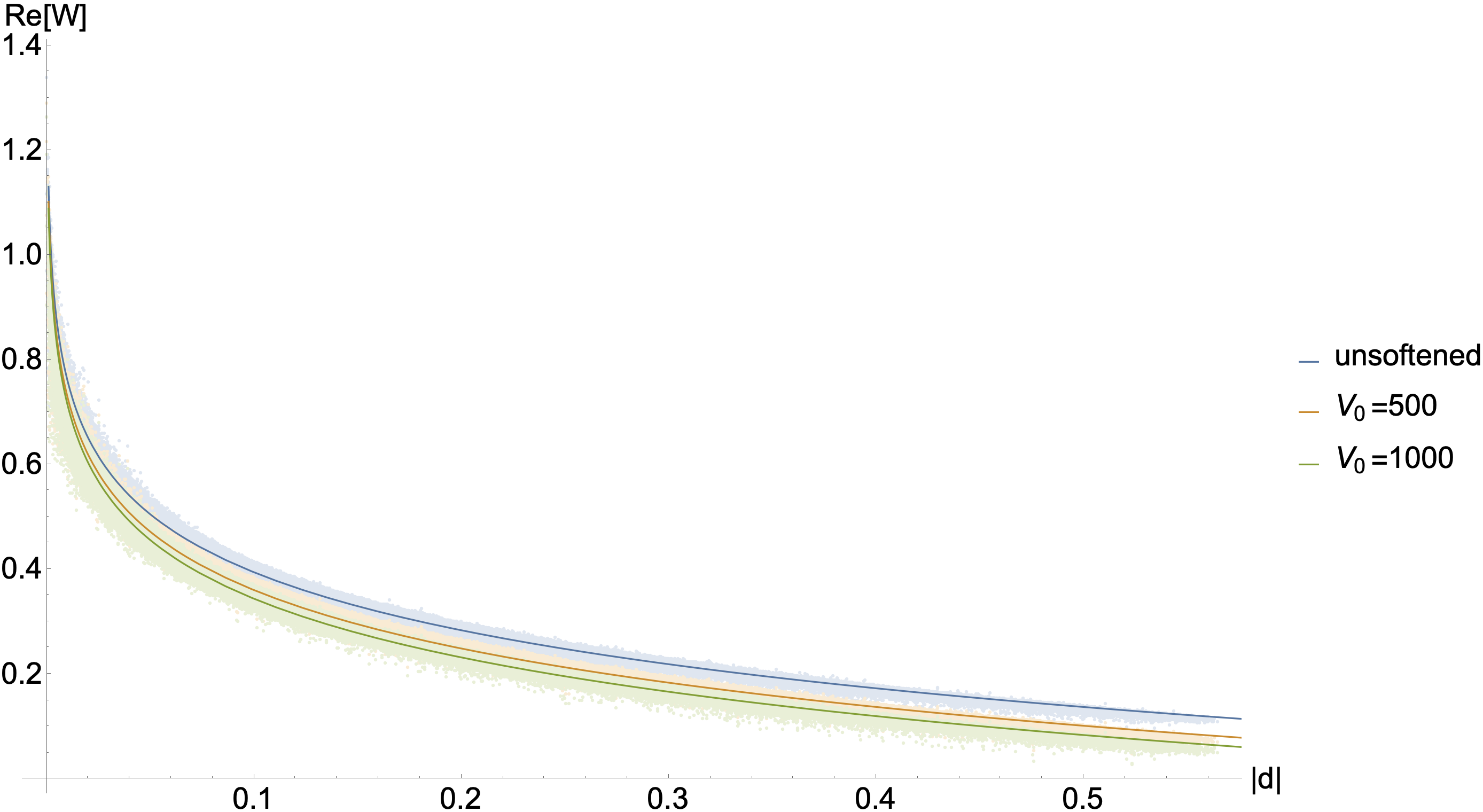}
\caption{The real part of the softened and unsoftened SJ Wightman function in a $20000$ element causal set versus geodesic distance. Best fit curves are shown, all of which are consistent with the scaling $-\frac{1}{2\pi}\ln|d|+\text{const}$, but with different constants.}
\label{fig:W_mid}
\end{figure}

\begin{figure}[h]
 \centering
 \begin{subfigure}{.75\textwidth}
     \includegraphics[width=\textwidth]{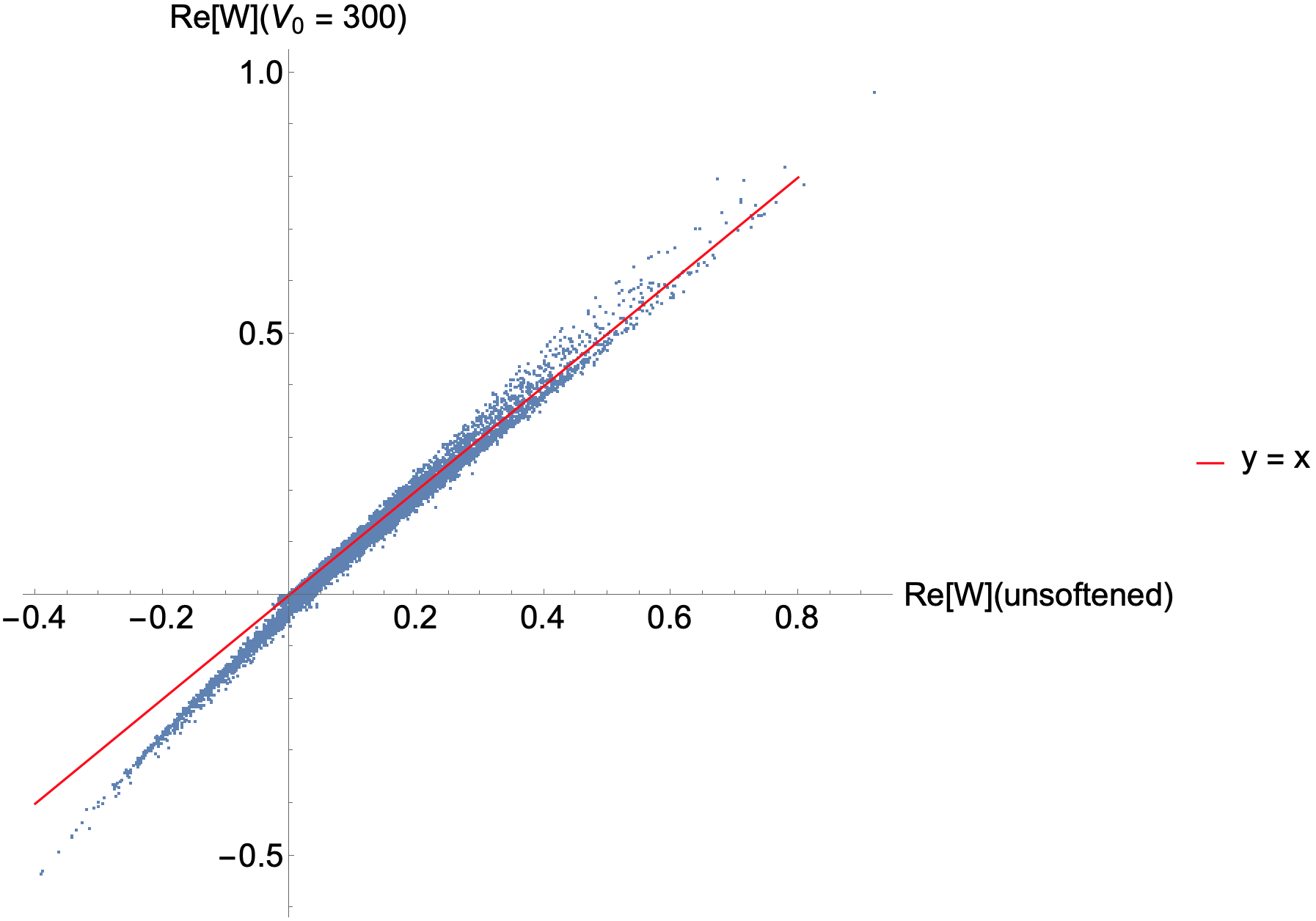}
     \caption{}
     \label{fig:corre300_nobound}
 \end{subfigure}
 \hfill
 \begin{subfigure}{.75\textwidth}
     \includegraphics[width=\textwidth]{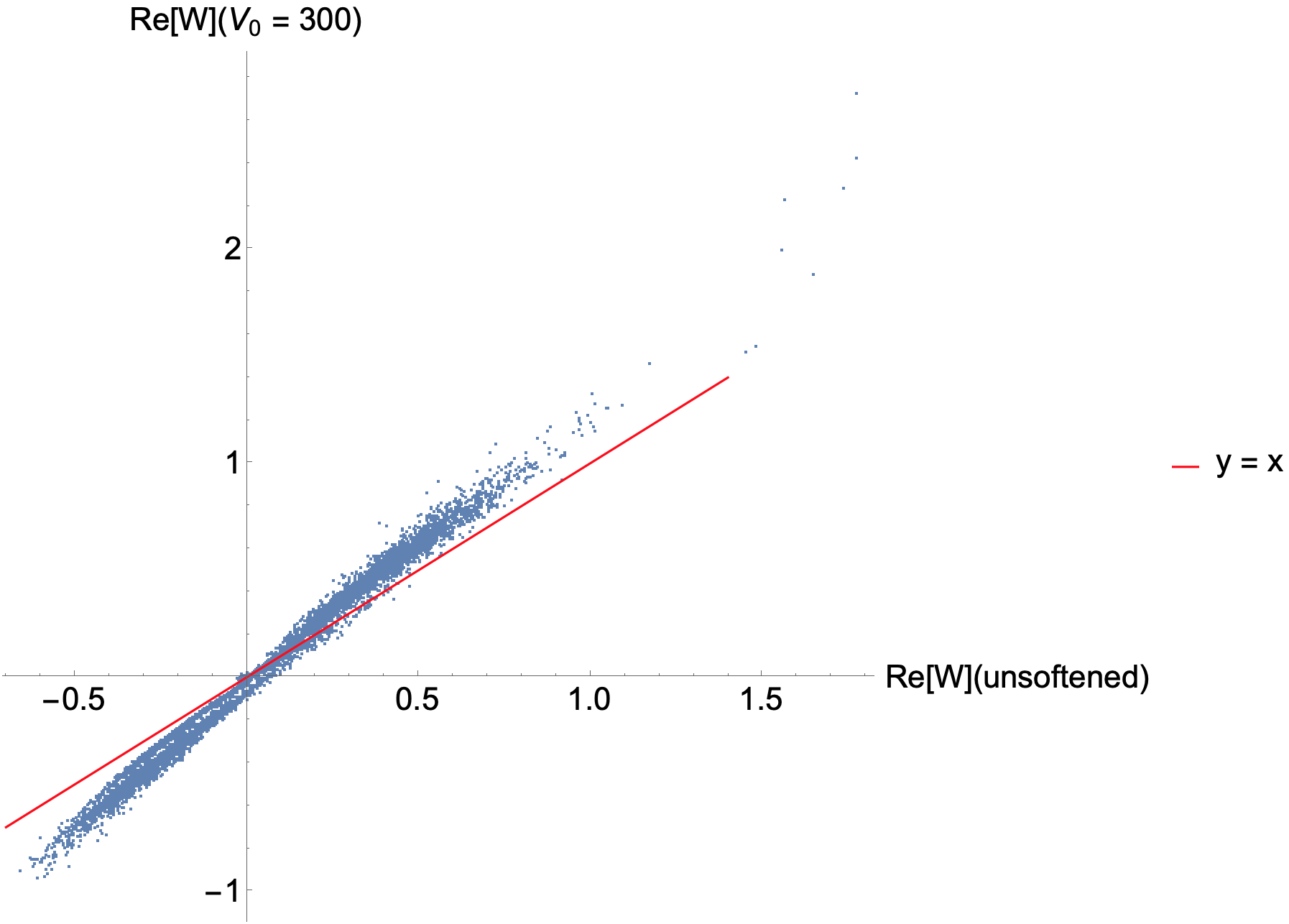}
     \caption{}
     \label{fig:corre300_bound}
 \end{subfigure}
 \caption{Correlation plot with 10$^4$ randomly-selected values of $\text{Re}[W_{\text{softened}}(x,x')]$ vs $\text{Re}[W_{\text{unsoftened}}(x,x')]$ (where the softened state has $V_0=300$ in \eqref{softenening_defs}), from points sampled from (a) 95$\%$ of the middle area ($|u|,|v|\leq0.0.975L$), and (b) 5$\%$ near the boundary ($|u|,|v|>0.0.975L$). For reference, the red line is the line with slope 1 and intercept 0, representing perfect positive correlation.}
 \label{fig:corre300}
\end{figure}

Figure \ref{fig:VRho} shows the relation between $V_0$ and the mean value of $\rho_i$ in a central subvolume of a $20000$ element causal set diamond with $L=$ 1. The central subvolume has $|t|\leq0.28 L$ and $|x|\leq0.28L$. In the same central region, we studied the behavior of the softened Wightman function versus the geodesic distance. The results for softenings with $V_0 = 500$ and $V_0 = 1000$ are shown in Figure \ref{fig:W_mid} along with the unsoftened case for comparison. All three Wightman functions are consistent with the expected Minkowski scaling behavior of $-\frac{1}{2\pi}\ln|d|+\text{const}$, but with different constants ($0.027, -0.010$, and $-0.028$ for the unsoftened, $V_0 = 500$ and $V_0 = 1000$ cases respectively). The value of the constant decreases as we increase $V_0$. This together with the $ - \frac{1}{2\pi}\textrm{ln} \left( \frac{\pi}{L} \right)$ term in \eqref{eqn:Wgeneral} suggest the interpretation that the softening may in some sense decrease the value of the IR cutoff. In Figure \ref{fig:corre300} correlation plots are shown of the softened versus unsoftened Wightman functions, for a case with $V_0=300$ and number of elements $N = 20000$. Separate correlation plots are shown for points sampled near the boundary and away from the boundary. We expect the softened and unsoftened Wightman functions to be more correlated away from the boundary, as the softening -- by design -- modifies the state near the boundary.  This is also what we see in the correlation plots, as the scatter plot for the central region overlaps more with the $y=x$ line, which indicates that the values more closely agree.

\subsection{Entanglement Entropy}
We now study the entanglement entropy of a Gaussian scalar field in the softened SJ state described above, in a 1+1D causal set diamond.  We will use the formulation of the entanglement entropy in terms of spacetime two-point correlation functions \cite{Sorkin_2012}. According to this formulation, the entanglement entropy is given by
\begin{equation}
S = \sum_{\sigma} \sigma \textrm{ln} |\sigma|,
\end{equation}
where the $\sigma$'s are obtained by solving the generalized eigenvalue problem
\begin{equation}
W v = i \sigma \Delta v.
\label{eq:geneig}
\end{equation}
$W$ is chosen to be the one associated to the SJ state and it is initially pure, meaning that it yields zero entropy according to the above formulation. When the generalized eigenvalue equation \eqref{eq:geneig} is solved, this must be done in the subregion (e.g. one side of an event horizon) whose entanglement with its causal complement we wish to quantify. This will no longer lead to a vanishing entropy.

 \begin{figure}[h]
\centering
\includegraphics[width=9cm]{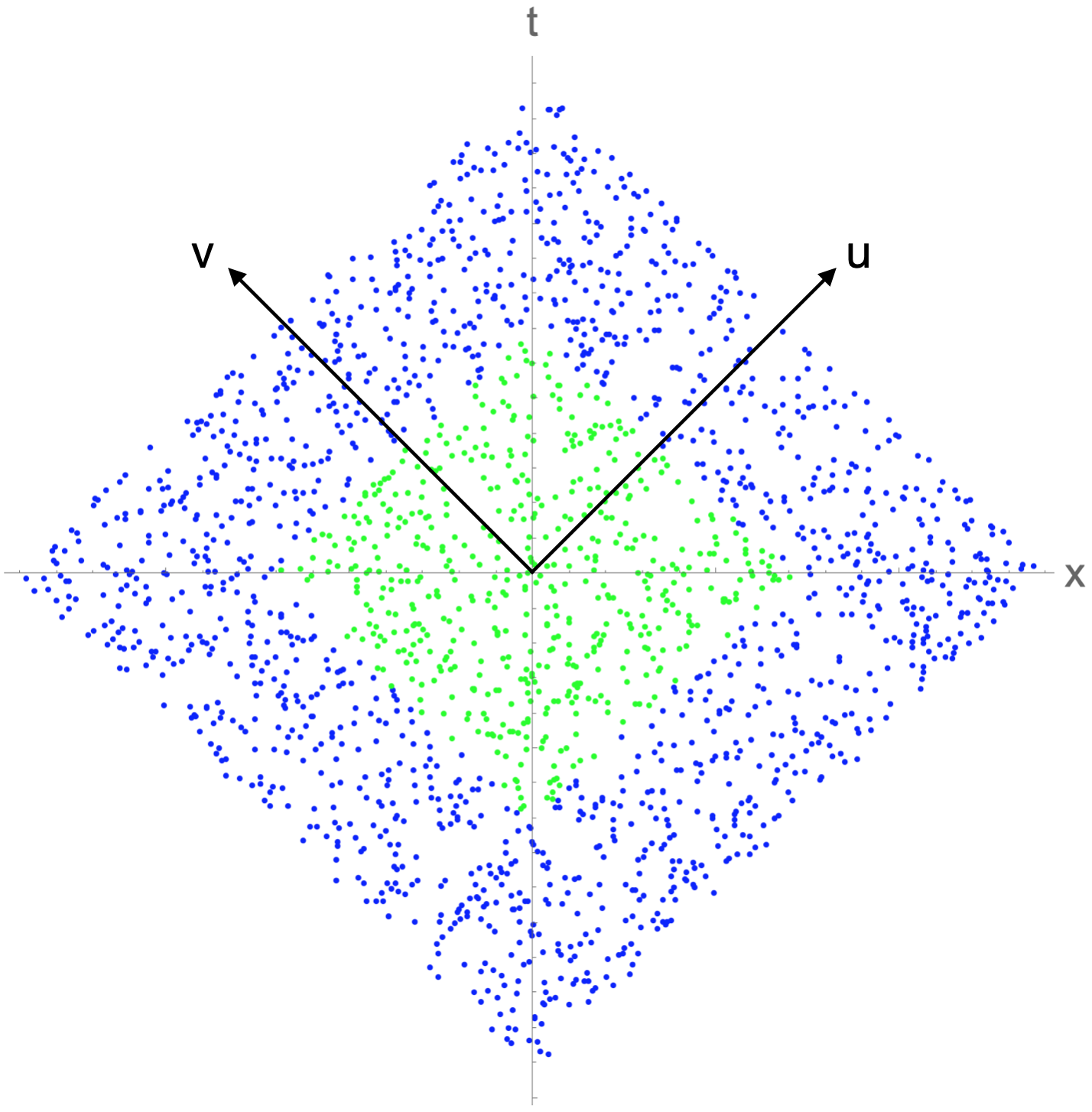}
\caption{An example causal set diamond with $N=2000$,  generated by sprinkling. The $(u,v)$ lightcone coordinate axes are shown. In this work, we study the entanglement entropy of a scalar field confined to the smaller (green) causal diamond within the larger (blue) one.}
\label{fig:diamond}
\end{figure}

\begin{figure}[h]
    \centering
    \includegraphics[width=.8\textwidth]{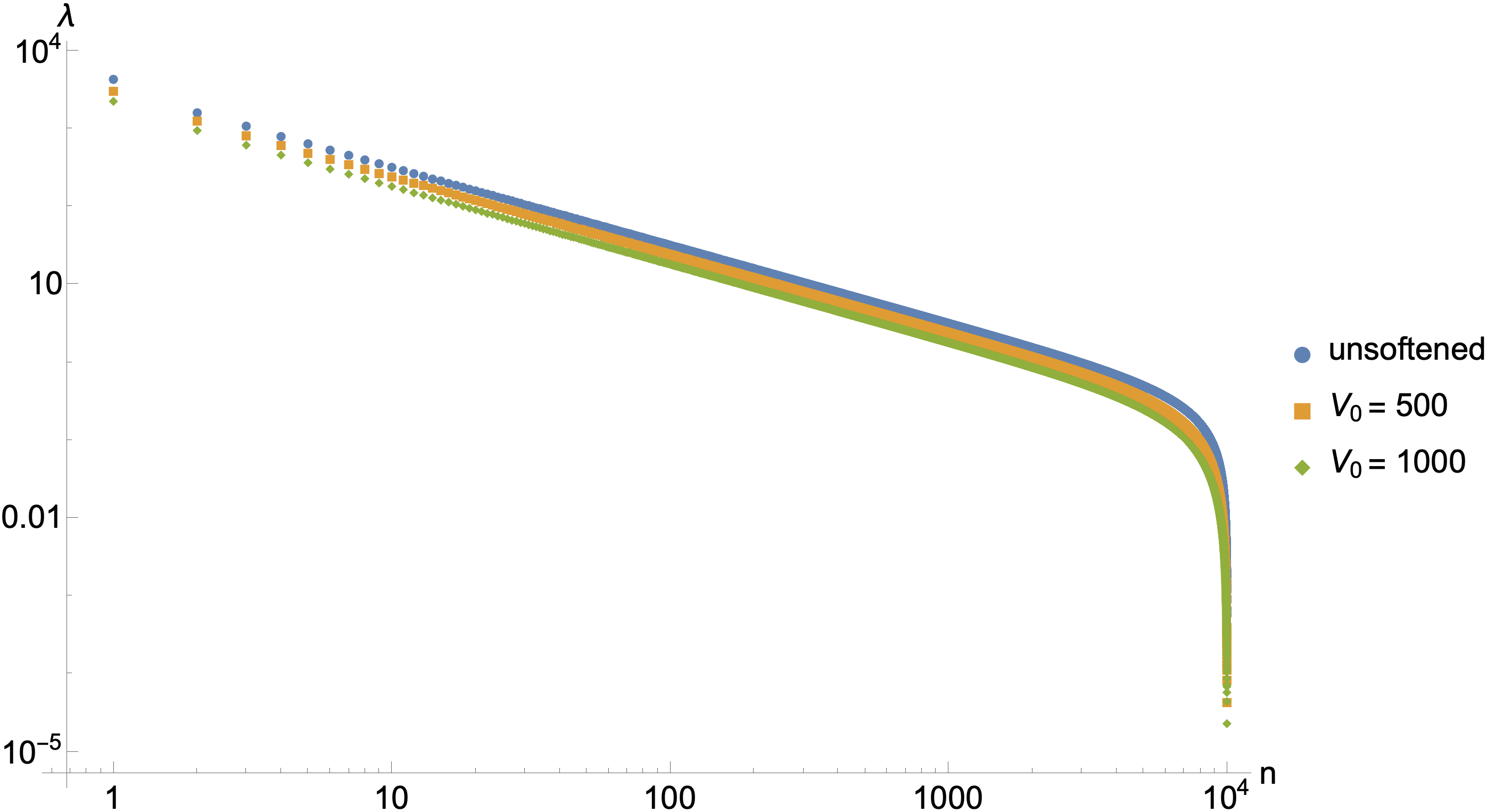}
    \caption{Positive eigenvalues of $i\Delta$ on a log-log scale, demonstrating both power law (large eigenvalues) and non-power law (small eigenvalues) regimes.}
    \label{fig:eigs}
\end{figure}

In \cite{Sorkin_2018} the entanglement entropy of a massless scalar field confined to a smaller causal diamond within a larger one was studied. Figure \ref{fig:diamond} shows the setup. These causal sets are generated by  sprinkling: randomly placing points within the manifold such that the number of elements within each generic volume statistically follows the Poisson distribution. In \cite{Sorkin_2018}, the scaling of the entanglement entropy with respect to the ultraviolet cutoff (discreteness scale in a causal set) was studied. Based on standard results, a spatial area-law scaling (logarithmic scaling of $S$ with $N$ in $1+1$D) was anticipated \cite{Calabrese_2009, Chandran_2016, Saravani_2014}, but instead a surprising spacetime volume-law (linear scaling of $S$ with $N$ in $1+1$D) was obtained \cite{Sorkin_2018}. The excess entropy was found to be related to numerous unexpected fluctuation-like components \cite{ee2}, in the small eigenvalue regime of the SJ eigendecomposition. These eigenvalues also have another marked difference from those expected to contribute to the entropy: they do not follow a power law when sorted from largest to smallest. An example spectrum of $i\Delta$ is shown in Figure \ref{fig:eigs} on a log-log scale. 
 As this unexpected result concerns the deep UV regime of the theory, and the Hadamard condition also concerns short distance behavior, we investigate whether these two things might be connected. We do this by repeating the calculation of the scaling of the entropy with respect to the UV cutoff, this time using the softened SJ state as our starting point (i.e. our initial pure state). 
 
 Seeing as the eigenvalues of $i\Delta$ in the softened case behave very similarly to the ones in the unsoftened case (as shown in Figure \ref{fig:eigs}), it seems unlikely that their entanglement entropy behavior would be any different either. Nevertheless there is still the possibility that the eigenfunctions and/or the scalings of the eigenvalues with the UV cutoff may create non-trivial differences. 
 
\begin{figure}[h]
     \centering
     \includegraphics[width=.85 \textwidth]{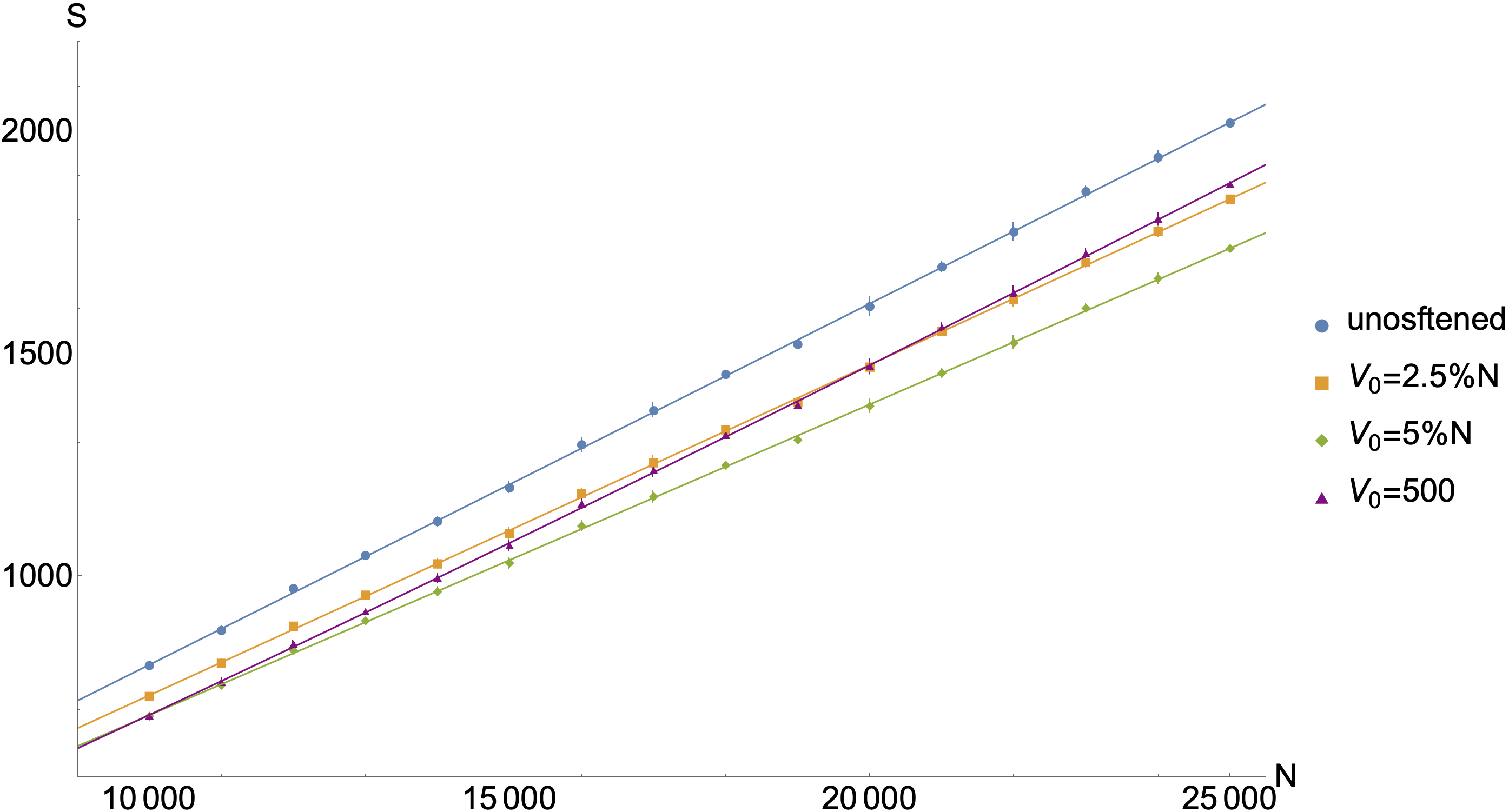}
     \caption{Entanglement entropy $S$ versus the number of elements $N$.}
     \label{fig:SN}
 \end{figure}

We investigated the entanglement entropy scalings with respect to the UV cutoff when the softened SJ state is used as the initial pure state. We sprinkled five causal set diamonds for each size from $N=10000$ to $N=25000$. We considered four cases of the  unsoftened, $V_0 = 2.5\% N$, $V_0 = 5\% N$ and $V_0 = 500$. The results are shown in Figure \ref{fig:SN} along with best fit lines which are consistent with a linear scaling of $S$ with $N$ and therefore a spacetime volume law. The magnitudes of the entropies in the softened case are smaller than those in the unsoftened case.

  Therefore,  using the softened SJ states as our initial pure states yields similar results for the entanglement entropy compared to using the unsoftened state as the initial pure state. We conclude from this that the excess entropy that leads to a volume law is  not connected to the non-Hadamard nature of the SJ state.
\section{Summary and Conclusions}
\label{sec:Summary}
With quantum gravity not within reach yet, we must take care in our assumptions about the ultimate UV completion of quantum field theory. What is a physical choice of vacuum state in arbitrary curved spacetimes? Hadamard states are a popular answer to this question. Hadamard states have a prescribed short-distance behavior, making them a minimal deformation away from the Minkowski vacuum.  They are an attractive class of state for reasons including that they yield a well-defined perturbation theory. The Sorkin-Johnston or SJ vacuum is an alternative and more recent answer to this question. It too has some advantages, such as being defined in a covariant and unique manner as well as being well defined in a causal set. However, the SJ state is not always Hadamard.

In this work, we studied the extent to which the SJ state in a $1+1$D causal diamond in Minkowski spacetime is Hadamard. In order to carry out our analysis, we extended what was known about the SJ state in the diamond, by approximating (both numerically and analytically) a subleading and previously little-studied  portion of the SJ Wightman function for this case. We then compared both the noncoincindence and the coincidence limit singularity structure of the SJ Wightman function with that of the Minkowski Wightman function, finding deviations that rendered the SJ state non-Hadamard at the boundary of the diamond. A similar conclusion was reached in \cite{Fewster_had1, Fewster_had2} through more rigorous arguments, but without studying the explicit nature of this departure from the Hadamard form.  We also found additional non-Hadamard nonlocal divergences on the boundary of the diamond in the leading part of the SJ state, when the points $x$ and $x'$ are on adjacent left or right edges of the boundary.

We also studied the SJ state in a causal set sprinkling of the causal diamond. Causal sets are thought to be the more fundamental underpinning of continuum spacetime. Additionally, they make computations and therefore the study of the SJ state and its modifications much easier to perform. We used the causal set calculation to study softenings of the SJ state, designed to make them Hadamard. As previously mentioned, while the softening does not strictly speaking make the SJ state Hadamard (as this is not meaningful in a discrete setting), it can be viewed as a discrete analogue of a Hadamard state.

One of the particular things we investigated was whether the softened SJ states can ameliorate some of the peculiar properties of entanglement entropy in causal set theory. In this context, there is an abundance of UV degrees of freedom (much greater than the number expected) that contribute to the entropy. This leads to a spacetime volume law scaling of the entropy with the discreteness scale instead of a spatial area law. We found that even when the starting point of the entanglement entropy calculation is the softened SJ state, although the magnitude of the entropy is smaller compared to the unsoftened case, the volume law scaling persists. Therefore we conclude that the entanglement entropy behavior and non-Hadamard property are likely not connected. Had we found that the softened SJ states led to the expected spatial area law, this would indicate a link between the extra entropy and the non-Hadamard nature of the SJ state; it would also, importantly, lend itself to one reason to prefer Hadamard states over non-Hadamard states in causal set theory. However, as it stands, we did not find evidence in favor of this. Furthermore, in recent work, interacting quantum field theories in causal set theory have been successfully constructed based on the SJ state \cite{Sorkin_2011, emma, Jubb:2023mlv}. These theories admit a well defined perturbation theory. Hence it may turn out that some reasons why Hadamard states are preferred would equally apply to SJ states, regardless of whether or not they are Hadamard.

\section*{Acknowledgments}
We thank Rafael Sorkin and Kasia Rejzner for helpful discussions. YY acknowledges financial support from Imperial College London through an Imperial College Research Fellowship grant.

\appendix
\section{Numerical Study of the Epsilon Term}
\label{app:NumericalEpsilon}
Approximating $\epsilon$ analytically allows one to directly evaluate the infinite sum with respect to $n$ in \eqref{eqn:epsilon_j}. However, this comes at the cost of loss in accuracy since our analytic treatment of $\epsilon^{(j)}$ consists of two Taylor series: \textit{S}$_2$ is expanded around $k_n^{(0)}$, and $\delta_n$ is approximated by expanding the transcendental equation. Below, we numerically compare $\epsilon^{(1)}$ with the exact error, in order to justify that the former is sufficient to capture the main short distance features of the state.
\subsection{$\epsilon(u,v;u,v)$}
\begin{figure}[h]
\centering
\includegraphics[width=13cm]{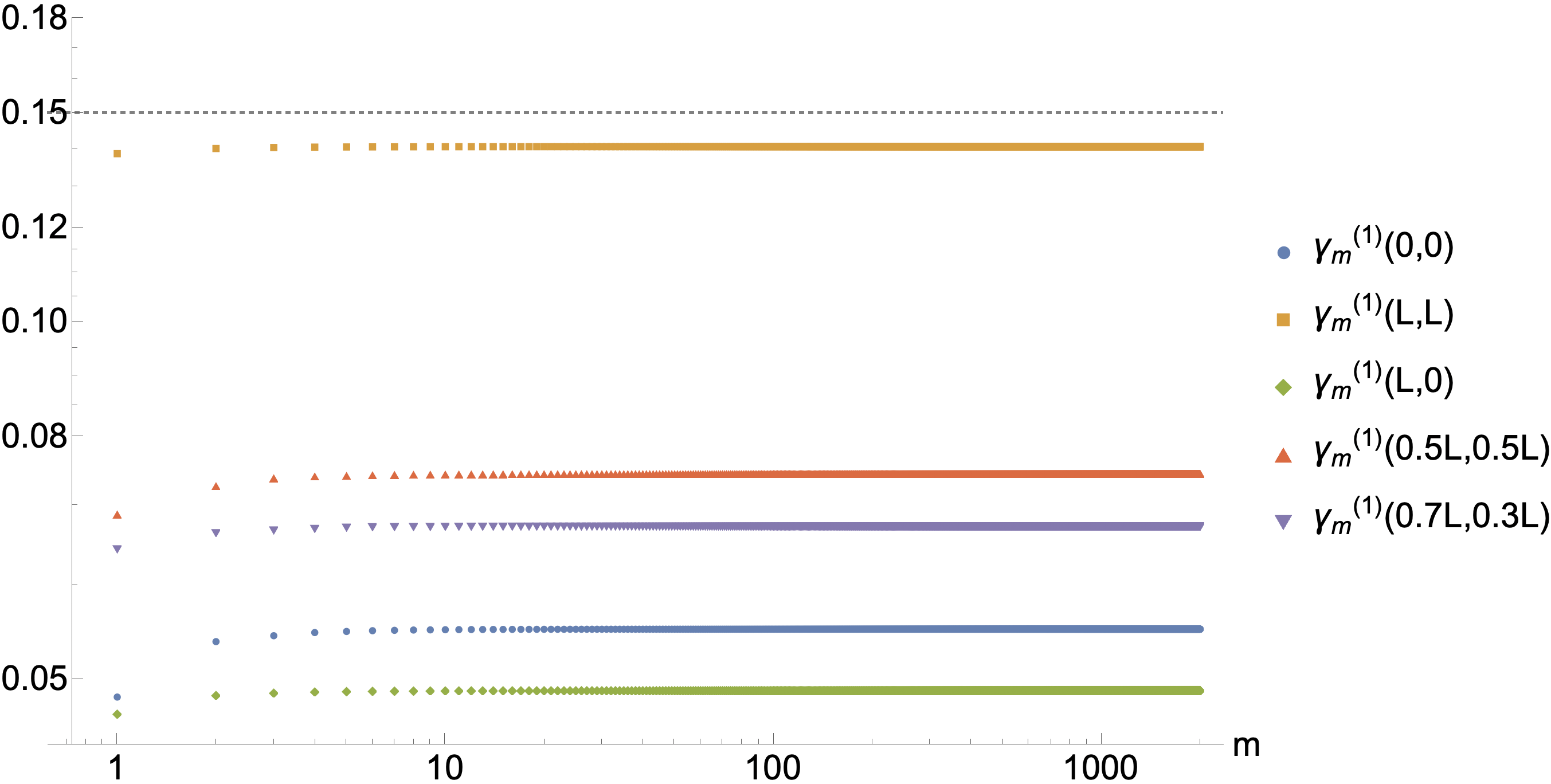}
\caption{$\gamma_m^{(1)}$ for different values of $m$ at several representative locations in the causal diamond.}
\label{fig:gamma}
\end{figure}

 We define $\gamma_m^{(1)}(u,v)$ as
\begin{equation}
\gamma_m^{(1)}(u,v) \equiv \sum_{n=1}^{m} \left(\epsilon_n(u,v;u,v) - \epsilon^{(1)}_n(u,v;u,v)\right),
\label{eqn:gamma_n}
\end{equation}
which is the cumulative sum of the difference between our analytic approximation $\epsilon_n^{(1)}(u,v)$ in \eqref{eqn:epsilon_j} and $\epsilon_n(u,v)$ numerically solved, up to $n=m$. Recall that $\epsilon_n$ is defined as \eqref{eqn:epsilon_exact_n}. The expression inside the summation in \eqref{eqn:gamma_n} is therefore what is left when the first two terms on the right hand side of \eqref{eqn:S2_1st_order} are moved to the left hand side.
 Note that in \eqref{eqn:gamma_n}, the coincidence limit ($u^\prime \rightarrow u$, $v^\prime \rightarrow v$) has been taken. We explicitly evaluate $\gamma_m^{(1)}$  for different values of $m$ up to $m=2000$ at several representative locations within the causal diamond. As can be seen in Figure \ref{fig:gamma}, $\gamma_m^{(1)}$ is bounded from above by some constant (gray dashed line). This suggests that we have not overlooked any divergences in our analytic approximations. Hence we conclude that $\epsilon^{(1)}$ approximates $\epsilon$ well, for the purposes of studying the (non)Hadamard form of the SJ state, and higher order terms in the series need not be considered.

\subsection{First Derivatives of $\epsilon(u,v;u,v)$}

\begin{figure}[h]
 \begin{subfigure}{0.41\textwidth}
     \includegraphics[width=\textwidth]{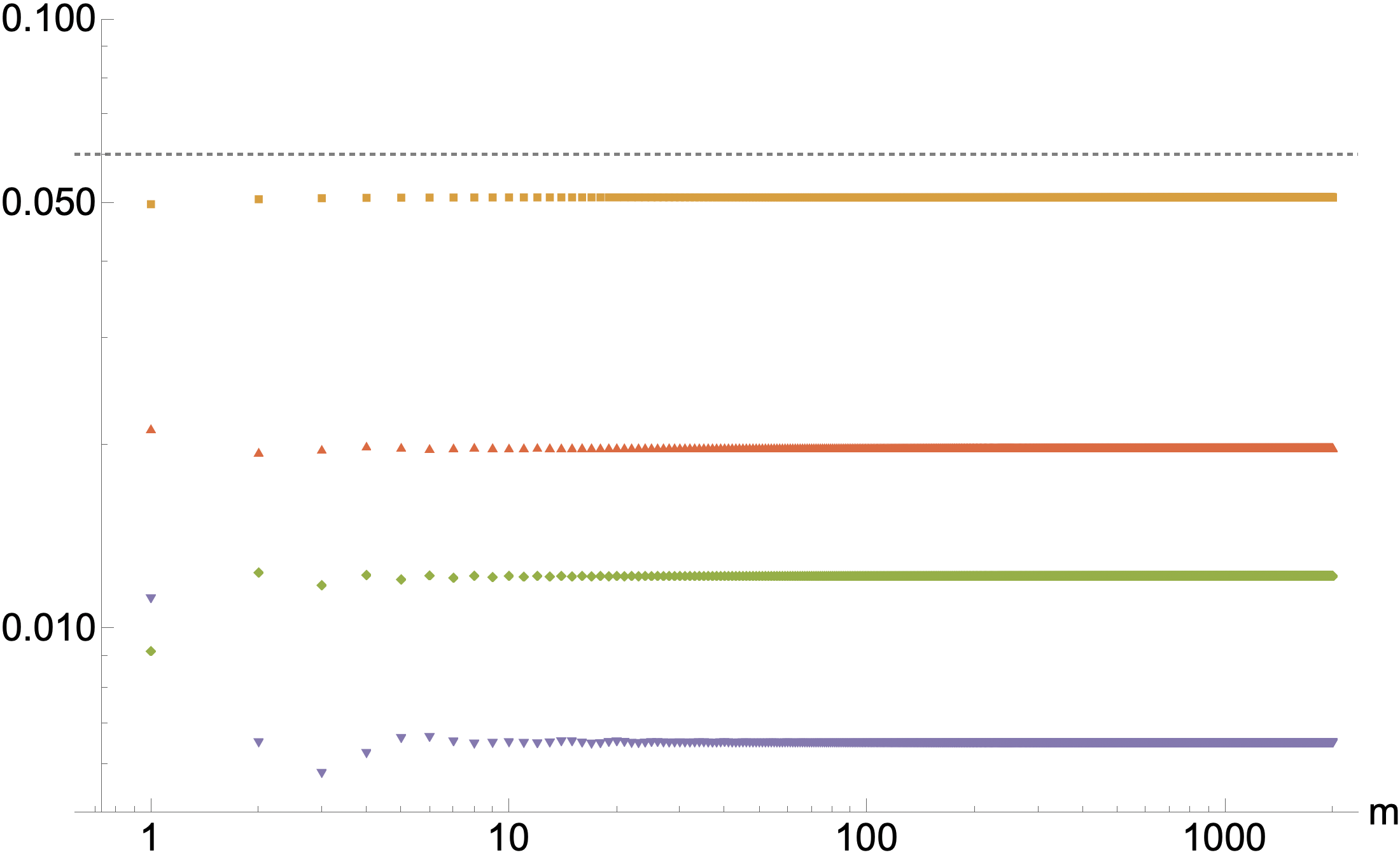}
     \caption{$Re[\gamma_{m,u}^{(1)}]$}
     \label{fig:gamma_u_Re}
 \end{subfigure}
 \begin{subfigure}{0.41\textwidth}
     \includegraphics[width=\textwidth]{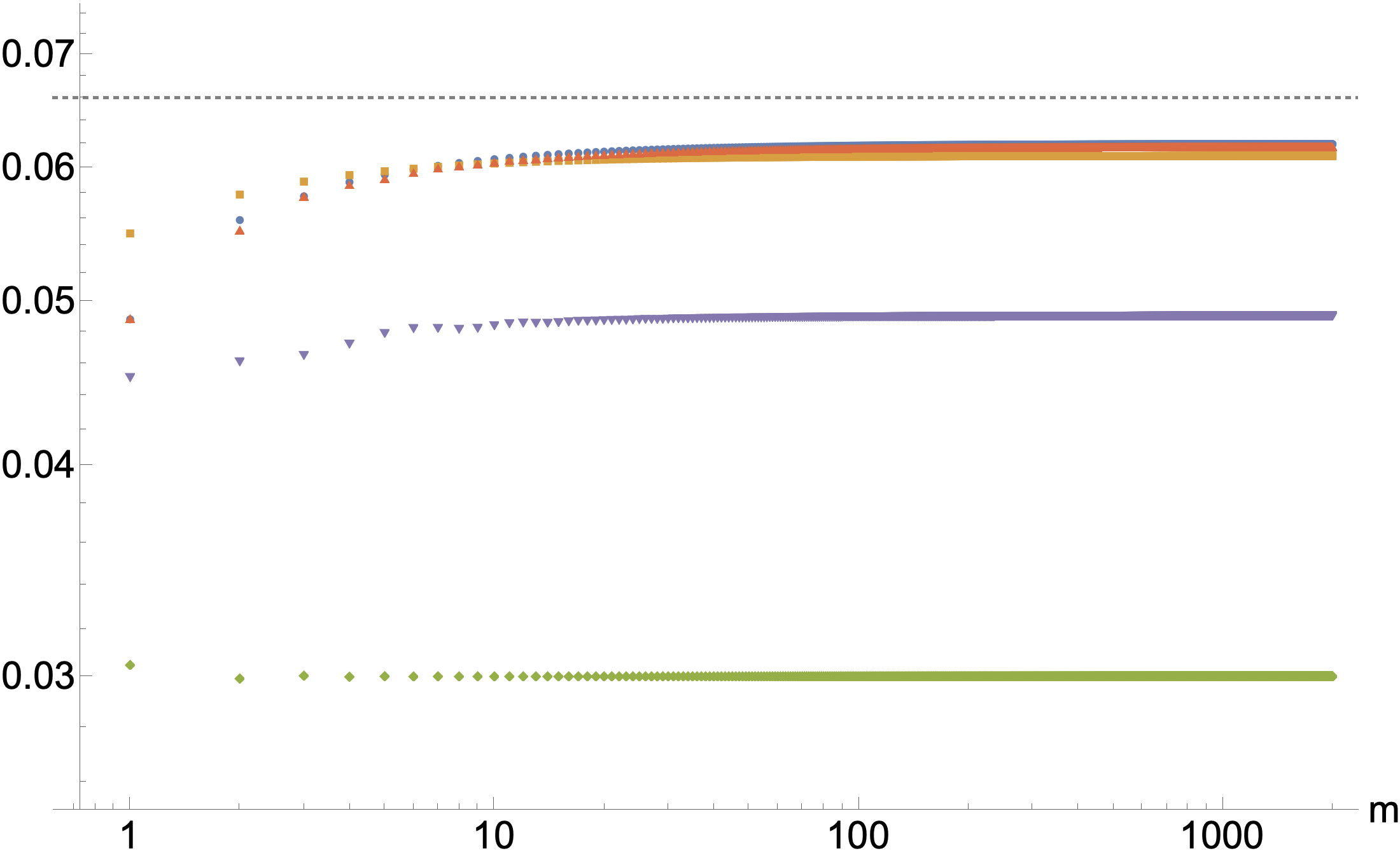}
     \caption{$Im[\gamma_{m,u}^{(1)}]$ }
     \label{fig:gamma_u_Im}
 \end{subfigure}
  \begin{subfigure}{0.15\textwidth}
     \includegraphics[width=\textwidth]{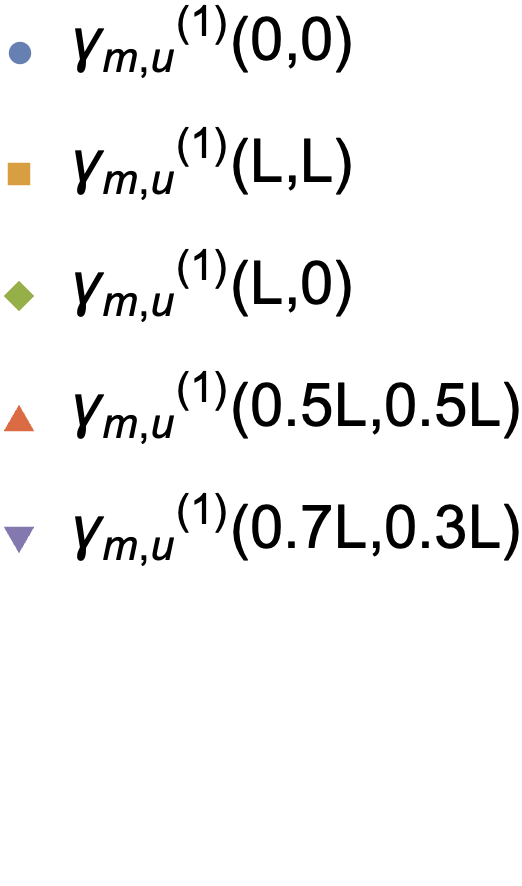}
 \end{subfigure}
 \caption{Real and imaginary parts of $\gamma_{m,u}^{(1)}$ for different values of $m$ at several representative locations in the causal diamond.}
 \label{fig:gamma_u}
\end{figure}

To see how well $\epsilon_{,u}^{(1)}$ approximates the exact error $\epsilon_{,u}$, we again define the cumulative sum up to the $m^{th}$ root of the difference between $\epsilon_{,u}$ and $\epsilon_{,u}^{(1)}$ at each $n$ to be
\begin{equation}
\gamma_{m,u}^{(1)}(u,v) \equiv \sum_{n=1}^{m} \left(\epsilon_{n,u}(u,v;u,v) - \epsilon^{(1)}_{n,u}(u,v;u,v)\right).
\end{equation}
\noindent
$\gamma_{m,u}^{(1)}$ are evaluated for different values of $m$ up to $m=2000$ at several representative locations inside the causal diamond; the results are shown in Figure \ref{fig:gamma_u}.  As can be seen in the figure, $\gamma_{m,u}^{(1)}$ is bounded from above by some constant (gray dashed line). This means that any divergences we found in our analytic approximations are real divergences and would not be cancelled away by further corrections. Therefore, we conclude that $\epsilon_{,u}^{(1)}$ alone is sufficient to approximate $\epsilon_{,u}$, for the purposes of studying its short distance behavior. 

\subsection{Second Derivatives of $\epsilon(u,v;u,v)$}

For studying the second derivatives, we follow a similar recipe as before and define the cumulative difference up to $n=m$ between $\epsilon_{,uu}$ and $\epsilon_{,uu}^{(j)}$ at each $n$ to be 
\begin{equation}
\gamma_{m,uu}^{(j)}(u,v) \equiv \sum_{n=1}^{m} \left(\epsilon_{n,uu}(u,v;u,v) - \sum_{i=1}^j \epsilon^{(j)}_{m,uu}(u,v;u,v)\right).
\end{equation}

For the second derivatives, in some cases it turns out to be necessary to go up to $j=2$ for the analytic approximations to capture the main short distance behavior accurately.

\bibliographystyle{iopart-num}
\bibliography{CS}

\end{document}